\documentclass[A4paper,12pt]{article}
\usepackage{amsmath,amsthm}
\usepackage{graphicx,psfrag,epsf}
\usepackage{enumerate}
\usepackage{natbib}
\usepackage{url} % not crucial - just used below for the URL 
 
\newcommand{\blind}{0}

\usepackage{xcolor}
\usepackage{ctable}
\usepackage{booktabs}
\usepackage{longtable}
\usepackage{multirow}

\usepackage{latexsym}
\usepackage{epsfig}
\usepackage{subfig}
\usepackage{lscape}
\usepackage{ulem}

\usepackage{rotating}

\newtheorem{theorem}{Theorem}
\newtheorem{lemma}{Lemma}

\begin{document}

\def\spacingset#1{\renewcommand{\baselinestretch}%
{#1}\small\normalsize} \spacingset{1}

%%%%%%%%%%%%%%%%%%%%%%%%%%%%%%%%%%%%%%%%%%%%%%%%%%%%%%%%%%%%%%%%%%%%%%%%%%%%%%

%\if1\blind
{
  \title{\bf Marginalized Frailty-Based Illness-Death Model: Application to the UK-Biobank Survival Data}
  \author{Malka Gorfine\thanks{
  		The authors gratefully acknowledge support from the NIH (R01CA189532), the U.S.-Israel Binational Science Foundation (BSF, 2016126), and the Israel Science Foundation (ISF, 1067/17) in carrying out this work
    }\hspace{.2cm}\\
   Department of Statistics and Operations Research, Tel Aviv University, Israel\\
    and \\
    Nir Keret\\
    Department of Statistics and Operations Research, Tel Aviv University, Israel\\
    and\\
    Asaf Ben Arie\\
   Department of Statistics and Operations Research, Tel Aviv University, Israel\\
   and\\
   David Zucker\\
	Departmet of Statistics, the Hebrew University, Jerusalem, Israel\\
	and\\
	Li Hsu\\
	Fred Hutchinson Cancer Research Center, Seattle, USA }
  \maketitle
}% \fi

\if0\blind
{
  \bigskip
  \bigskip
  \bigskip
  \begin{center}
    {\LARGE\bf }
\end{center}
  \medskip
} \fi

\bigskip
\begin{abstract}
The UK Biobank is a large-scale health resource comprising genetic, environmental and medical information on approximately 500,000 volunteer participants in the UK, recruited at ages 40--69 during the years 2006--2010. The project monitors the health and well-being of its participants. This work  demonstrates how these data can be used to estimate in a semi-parametric fashion the effects of genetic and environmental risk factors on the hazard functions of various diseases, such as colorectal cancer. An illness-death model is adopted, which inherently is a semi-competing risks model, since death can censor the disease, but not vice versa. Using a shared-frailty approach to account for the dependence between time to disease diagnosis and time to death, we provide a new illness-death model that assumes Cox models for the marginal hazard functions. The recruitment procedure used in this study introduces delayed entry to the data. An additional challenge arising from the recruitment procedure is that information coming from both prevalent and incident cases must be aggregated.
Lastly, we do not observe any deaths prior to the minimal recruitment age, 40.
In this work we provide an estimation procedure for our new illness-death model that overcomes all the above challenges. 
\end{abstract}

\noindent%
\vfill

\newpage
\spacingset{1.5} % DON'T change the spacing!
\section{Introduction}
\label{sec:intro}
The UK Biobank (UKB) is a large-scale health resource comprising genetic, medical and environmental information on approximately 500,000 volunteer participants in the UK, recruited at ages 40--69 during the years 2006--2010. The project monitors the health and well-being of its participants, thus providing a strong incentive for joining. The participants have undergone various measurements, provided blood, urine and saliva samples for future analysis, and have also provided detailed information about themselves.  In using these data to study a given disease, the participants can be classified into three groups: those already diagnosed with the disease at the time of recuitment ("prevalent" cases), those in whom the disease is diagosed during follow-up ("incident" cases), and those not diagnosed with the disesase as of the end of follow-up.

One concern in such a design is delayed entry, since subjects need to live at least up to the minimum recruitment age in order to participate in the study. Moreover, the existence of prevalent cases in the data requires special attention due to two reasons: (i) The delayed-entry correction for those observations must be different to that for the incident cases. (ii) The data are subject to recall bias -- when participants are asked to provide information regarding their status at  the time of disease diagnosis, it is likely that some information will be reported inaccurately, especially if a long time has passed since then. Death before the disease constitutes a competing risk to many studied phenotypes, and while death in general can censor the disease, the disease cannot censor death, hence these settings are termed ``semi-competing risks''. 
The purpose of this work is to provide appropriate model and estimation procedure for estimating the survival distribution of a certain disease, such as colorectal cancer, that properly accommodate semi-competing risks and biased sampling due to delayed entry.

In this work we focus our attention on colorectal cancer data. Out of 484,918 participants with available genetic and environmental data, there is an overall number of 5,131 colorectal cancer cases in the UKB, of which 2,339 are prevalent and 2,792 are incident. The number of deaths before having the disease is 12,767, and the number of deaths after colorectal cancer diagnosis is 1,040. Over the years the number of incident cases will grow, while the number of prevalent cases will no longer change as recruitment is already complete.

Under semi-competing risks settings, three stochastic process are typically studied: the time until disease onset, time until death free of the disease, and time until death after disease onset. These three processes are sometimes measured on a sojourn time scale, namely, the disease-death process is not expressed as age of death after disease, but rather as the amount of time spent in diseased state until death. Since we wish to use a shared random effect to describe the dependence between the processes, it is easier to work only with the age-scale and avoid situations with negative dependence (e.g. between age at diagnosis and the sojourn time in diseased state until death).

\citet{fine2001semi} considered a gamma frailty model defined on the upper wedge of the joint distribution of the events, and supplied a consistent estimator for the parameter of the gamma distribution, but did not incorporate covariates. \citet{xu2010statistical} proposed gamma-frailty illness-death conditional and marginal models, such that the conditional (on the frailty variate) model is a Cox-type model. If the marginal model is of primary interest,  interpretation of the regression effects might become cumbersome as the marginal distribution does not take a simple form and also includes the frailty distribution parameter. \citet{chen2012maximum} assumed a semi-parametric transformation model for the marginal regressions and a copula model for the joint distribution. However, it was assumed that occurrence of the non-terminal event does not alter the distribution of the terminal event, which is unrealistic in most illness-death scenarios. Extending the copula model in order to account for the change of distribution is not straightforward. \citet{vakulenko2016comparing} considered an illness-death model with delayed entry and right-censored data, under a fully parametric regression framework. However, no dependence structure between the disease and death times was assumed beyond the observed covariates. \citet{vakulenko2017nonparametric} describe a non-parametric inverse probability weighting (IPW) approach for estimating the joint distribution of disease and death times, subject to right censoring and delayed entry. This approach does not incorporate covariates. In addition, under the sampling scheme present in the UKB data, an IPW approach is inapplicable (as will be explained in Section 3.3). \citet{zhou2016semiparametric} described a simple pseudo-likelihood approach with copulas, aimed at estimating the marginal survival functions and the association parameter of the copula, but did not account for delayed entry and did not incorporate covariates. The approach of pseudo-values was presented by \citet{andersen2007regression},  in order to directly estimate the covariate effects on the state probabilities, using a generalized estimating equations procedure. This approach requires the user to predetermine the time grid for the state probabilities. In addition, calculation of the pseudo-values for a dataset as big as the UKB poses a big computational burden. 

In illness-death models, age at death after disease diagnosis is left truncated by the age at diagnosis. In most applications it is unrealistic to assume that the observed covariates contain all the sources of dependence between age at diagnosis and age at death. Limited literature exists on Cox regression with dependent left-truncation and right-censored data. A complete review can be found in \cite{shen2017semiparametric}. In Section 4.2 we analyze the UKB data also by including age at diagnosis as a covariate in the model of age at death after diagnosis, in the spirit of  \cite{mackenzie2012survival} and \cite{shen2017semiparametric}. The troubling meaning of these analyses will be discussed.

None of the aforementioned approaches provides a satisfactory framework for the analysis of the UKB data. We are seeking a model that can accommodate delayed entry and a dependence structure between the three described processes. In addition, we want a model that is easy to interpret at the population level and can incorporate risk factors as covariates. Lastly, we would advise against using a fully parametric model, as these models require more assumptions and are typically less robust, but rather consider a semi-parametric framework. %, which will be shown to be relatively less sensitive to model misspecifications.

The novelty of this work consists of several aspects: (i) Formulation of a new frailty-based illness-death model with Cox-type marginalized hazards. Under random sampling and right-censored data, the model parameters are consistently estimated. (ii) Adjusting the proposed  estimators to accommodate delayed entry and the presence of both prevalent and incident cases, such as in the UKB data. R code for carrying out the data analysis and the
simulations reported in this paper is available
at the following \url{Github} site: \url{https://github.com/nirkeret/frailty-LTRC}.
 
The rest of the paper is organized as follows. Section~2 describes our proposed frailty-based illness-death model, and the pseudo-likelihood approach for estimating the regression coefficients and the baseline hazard functions for simple cohort studies with no delayed entry. 
The estimation procedure for delayed-entry data is outlined in Section~3.  In Section~4 we present the analysis of colorectal cancer data in the UKB. Section~5 summarizes simulation results, with and without delayed entry. Final remarks are presented in Section~6.

\section{Methods: right-censored data, no delayed entry}
\subsection{Models}
Two types of hazard functions are considered: a conditional hazard given the unobserved frailty variate and the observed time-independent covariates, and a marginalized hazard function with respect to the frailty variate, namely, the hazard given the time-independent covariates. The main goal is estimating the illness-death marginalized hazards and survival functions. \citet{xu2010statistical} defined a frailty-based illness death model such that the conditional hazards follow Cox-type models multiplied by a frailty variate, while the marginalized hazards are functions of the frailty-distribution parameter  \citep[][Eq.'s 19--20]{xu2010statistical}. In contrast, we adopt the approach of \citet{glidden1999semiparametric}, and reformulate the frailty model so that the marginalized hazard functions obey a specified model, such as the Cox model, that is free of the frailty-distribution parameter. In the context of this work, it is preferable that the marginal model be free of the frailty parameter, as its interpretation as a model corresponding to a randomly selected individual from the population is facilitated this way. The frailty distribution parameter quantifies the degree of dependence between the different processes within the same person, so when it is present in a marginal model it obscures the interpretation of the regression coefficients.

\begin{figure}[!t]
	\begin{center}
		\includegraphics[width=0.5\linewidth]{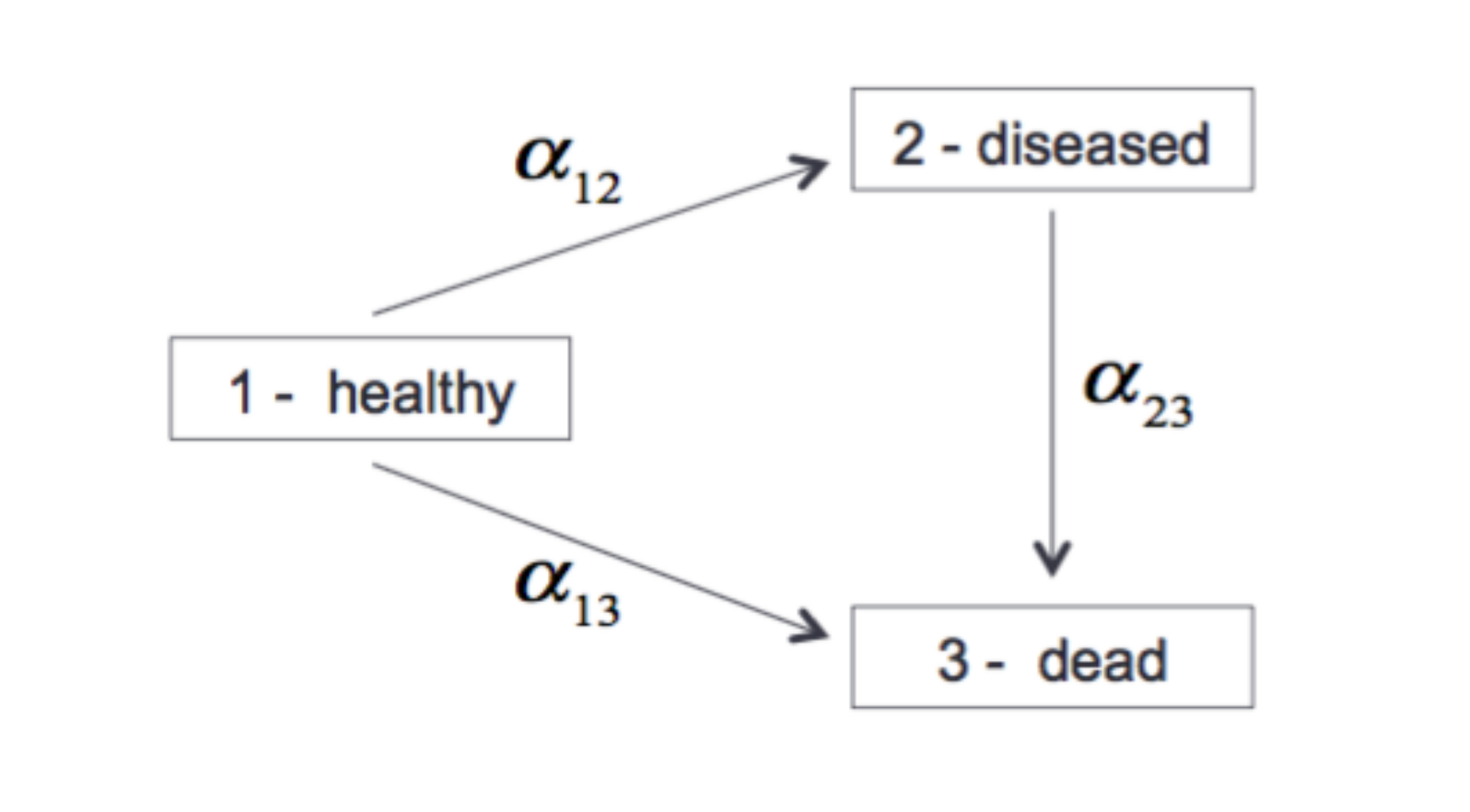}
		\caption{Illness-death process.\label{illness_death}}
	\end{center}
\end{figure}

Let $T_{1}$ and $T_{2}$  be age at diagnosis and age at death, respectively. Denote the unobserved frailty by a random variable $\omega >0$ with a known cumulative distribution function $F$ and an unknown parameter $\theta$. Let $Z$ be a vector of time-independent covariates. Based on the notation of Fig.~\ref{illness_death}, let the conditional hazards of transition from state 1 to either state $k=2$ or 3, given $(Z,\omega)$, be 
$$
h^c_{1k}(t|Z,\omega) = \lim_{\Delta \searrow 0} \frac{1}{\Delta} 
\Pr(t \leq T_{k-1} < t +\Delta| T_{1} \geq t, T_{2} \geq t,Z,\omega) = \omega  \alpha_{1k}(t|Z)  \;\;\; t>0 
\;\;\; k=2,3 \, .
$$ 
Let the conditional hazard function of leaving state 2, given $(Z,\omega)$, and given the transition to state 2 occurred at age $t_1$ be defined by
$$
h^c_{23}(t|t_1,Z,\omega) = \lim_{\Delta \searrow 0} \frac{1}{\Delta} 
\Pr(t \leq T_{2} < t +\Delta| T_{1}=t_1, T_{2} \geq t,Z,\omega)= \omega \alpha_{23}(t|Z)  \;\;\; t>t_1>0 \, .
$$ 
The non-negative functions $\alpha_{jk}$, $jk=12,13,23$, will be determined by the distribution of the frailty $\omega$ and the marginalized hazards presented below. The frailty distribution $F$ should be chosen such that the hazard models are identifiable.

The corresponding Cox marginalized hazards are defined by
$$
h_{1k}(t|Z) = \lim_{\Delta \searrow 0} \frac{1}{\Delta} 
\Pr(t \leq T_{k-1} < t +\Delta| T_{1} \geq t,T_{2} \geq t,Z) =  h_{01k}(t) \exp(\gamma_{1k}^T Z) \;\;\; t>0  \;\;\;  k=2,3 
\,\, ,
$$
and
$$
h_{23}(t|t_1,Z) = \lim_{\Delta \searrow 0} \frac{1}{\Delta} 
\Pr(t \leq T_2 < t +\Delta| T_{1}=t_1, T_{2}>t, Z) = h_{023}(t) \exp(\gamma_{23}^T Z) \;\;\; t>t_1>0 \, ,
$$
where $\gamma_{jk}$ and $h_{0jk}$, $jk=12,13,23$, are the regression coefficient vectors and unspecified baseline hazard functions, respectively. In $h_{23}$, disease onset time $t_1$ is not included in the vector of covariates, but instead the dependence between the time to disease onset and the time from disease onset to death is captured by the frailty parameter $\omega$.
Our goal is estimating the regression coefficients  $\gamma_{jk}$, the hazards, $h_{0jk}$, $jk=12,13,23$, and the dependence parameter $\theta$.  

If one is interested in estimating only $\gamma_{jk}$ and $h_{0jk}$ with $jk=12,13$, the standard partial likelihood approach can be applied, as in standard applications of Cox models with competing risks \citep[][Ch.~8]{kalbfleisch2011statistical}. Estimation of $\gamma_{23}$ and $h_{023}$ could be more involved since  age of death is left-truncated by the age at disease diagnosis. Under the (unreasonable) assumption that $T_1$ and $T_2$ are conditionally independent given $Z$, $\gamma_{23}$ and $h_{023}$ can be easily estimated using a standard partial likelihood approach and a Breslow estimator, with the usual risk-set correction for left-truncated data (see Section S8 of Supplementary Material). However, in most applications one cannot observe all the environmental and genetic effects that fully explain the dependence between $T_1$ and $T_2$, and the above conditional independence assumption is violated. The standard approach yields biased estimators of $\gamma_{23}$ and $h_{023}$, as will be demonstrated in the simulation study (Seciton 4).  Instead, we let $\omega$ represent the unobserved residual dependence, and assume that $T_1$ and $T_2$ are independent given $(Z,\omega)$. By this, we will be able to circumvent the dependent left-truncation problem. We provide a unified estimation procedure for all the parameters of interest, including the dependence parameter. 

As a final step of our new illness-death model presentation, we derive the relationship between $\alpha_{jk}$ and $h_{0jk}$, $jk=12,13,23$, for a given frailty distribution with differentiable inverse Laplace transform. Denote by $\phi(s)=E\{ \exp(-s \omega)\}$ the Laplace transform of $\omega$, by $\phi^{(q)}(s)$ its $q$th derivative with respect to $s$, by $\psi(s)$ the inverse Laplace transform, by $\psi^{(q)}(s)$ its  $q$th derivative, and by $\xi(s)$ the inverse of $-\phi^{(1)}(s)$. Also, let $H_{1.}(t|Z) = H_{12}(t|Z)+H_{13}(t|Z)$, $H_{0jk}(t) = \int_0^t h_{0jk}(u) du$, and $H_{jk}(t|Z) = \int_0^t h_{jk}(u|Z) du$, $jk=12,13,23$.

\begin{lemma}\label{lemma1} 
For $t>0$, the relationships between $\alpha_{jk}$ and $h_{jk}$, $jk=12,13,23$, are given by
\begin{eqnarray*}\label{eq:alpha12n}
\alpha_{1k}(t|Z) =  - h_{01k}(t) \exp(\gamma_{1k}^T Z) \psi^{(1)}[\exp\{-H_{1.}(t|Z)\}]\exp\{-H_{1.}(t|Z)\} \,\,\, ,
\,\, 
k = 2,3 \, \, ,
\end{eqnarray*}
and
\begin{eqnarray*}\label{eq:alpha23n}
\alpha_{23}(t|Z) &=& -\xi^{(1)} [\exp\{-H_{023}(t) \exp(\gamma_{23}^T Z) \} ] \exp\{-H_{023}(t) \exp(\gamma_{23}^T Z) \} \nonumber \\
&& \exp(\gamma_{23}^T Z) h_{023}(t) \, . 
\end{eqnarray*}
\end{lemma}
The proof of Lemma \ref{lemma1} is provided in Section S1 of the Supplementary Material.

Generally, a  marginalized proportional-hazards model does not yield a conditional proportional hazards model. Importantly, $\alpha_{jk}$ are of the form  
\begin{equation}\label{eq:alphastart}
\alpha_{jk}(t|Z)=h_{0jk}(t) \alpha^*_{jk}(t|Z) \,\,\, jk=12,13,23 \, ,
\end{equation}
where $\alpha^*_{jk}$ can be derived from the LT of the frailty distribution. For example, under the gamma-frailty model with expectation 1 and variance $\theta$, $\phi(s) =(1+\theta s)^{-1/\theta}$, and thus
\begin{equation*}
\alpha^*_{1k}(t|Z) = \exp(\gamma_{1k}^T Z)  \exp\{\theta H_{1.}(t|Z)\}  \, , \,\,\,  k=2,3 \, ,
\end{equation*}
and
\begin{equation*}
\alpha^*_{23}(t|Z) = \exp(\gamma_{23}^T Z)\exp\{ H_{23}(t|Z) \theta/(1+\theta) \}/(1+\theta)
 \,  .
\end{equation*}
As will be shown in the following section, the representation provided by Eq.~(\ref{eq:alphastart})  plays an important role in our proposed estimation procedure.

\subsection{Estimation Procedure}
Suppose there are $n$ independent subjects. For the $i$th subject, $i=1,\dots,n$, denote by $C_{i}$ the censoring time. Let $V_{i}=T_{1i} \wedge T_{2i} \wedge C_{i}$, $\delta_{1i}=I\{ T_{1i} \leq (T_{2i} \wedge C_{i}) \}$, so that  $\delta_{1i}$ equals 1 if the subject was observed to have the disease before being censored or dying. Also let 
$\delta_{2i}=I\{ T_{2i} \leq (T_{1i} \wedge C_{i})\}$, so that  $\delta_{2i}$ equals 1 if the subject died before having the disease or being censored. Denote by 
$W_{i}= \delta_{1i} (T_{2i} \wedge C_{i}) $ the age at death or age at censoring after having the disease,  and $\delta_{3i}=\delta_{1i} I(T_{2i} \leq C_{i})$, which equals 1 if death after the disease was observed.  Then, the observed data  consists of
$(V_{i},\delta_{1i},\delta_{2i},W_{i},\delta_{3i},Z_i)$, $i=1,\ldots,n$. The unobserved frailties $\omega_i$, $i=1,\ldots,n$, are assumed to be independent random variables with cumulative distribution function $F$, unknown parameter $\theta$ and LT $\phi$ such that $\psi^{(1)}$ and $\xi^{(1)}$ exist. 

Let $\gamma=(\gamma_{12}^T,\gamma_{13}^T,\gamma_{23}^T)^T$ and $H_0=(H_{012},H_{013},H_{023})$.  
The regression coefficients $\gamma$ will be estimated by maximizing a pseudo likelihood, while the cumulative hazard functions $H_0$ will be estimated with Breslow-type estimators. Since the likelihood contains $H_0$, and the Breslow-type estimator in turn requires $\gamma$ and $\theta$, a circular dependence is created, which calls for an iterative algorithm. The proposed estimation procedure is an extension of \citet{gorfine2006prospective} for the standard shared-frailty models of  correlated failure times.  A discussion that compares the proposed method and that of \citet{gorfine2006prospective} is provided in Section S2 of the Supplementary Material.
  
It is assumed that conditional on $Z_i$ and $\omega_i$, the censoring times are independent of the failure times and non-informative for $\omega_i$ and all the other parameters in the models. In addition, the frailty variate $\omega_i$ is assumed to be independent of $Z_i$. Then, the likelihood function is proportional to $L(\gamma,\theta,H_0) = \prod_{i=1}^n L_i$, where
\begin{eqnarray} \label{eq:likelihood}
L_i
&=&  \int f(V_i,\delta_{1i},\delta_{2i},W_i,\delta_{3i}|Z_i,\omega)dF(\omega)  \nonumber \\
&\propto &
\{h_{012}(V_i) \alpha^*_{12}(V_i|Z_{i}) \}^{\delta_{1i}}
\{h_{013}(V_i) \alpha^*_{13}(V_i|Z_{i})\}^{\delta_{2i}}
\{h_{023}(W_i) \alpha^*_{23}(W_i|Z_{i})\}^{\delta_{3i}} \nonumber \\
& & 
(-1)^{\delta_{.i}} \phi^{(\delta_{.i})}(s_i) \, ,
\end{eqnarray}
$\delta_{.i}=\sum_{j=1}^3 \delta_{ji}$, $A_{jk}(t|Z)=\int_0^t \alpha_{jk}(u|Z)du$, $jk=12,13,23$, and
\begin{eqnarray}\label{eq:si}
s_i=A_{12}(V_i|Z_{i})+A_{13}(V_i|Z_{i})+\delta_{1i}A_{23}(W_i|Z_{i})-\delta_{1i}A_{23}(V_i|Z_{i}) \, .
\end{eqnarray}
A detailed explanation of Eq.~(\ref{eq:likelihood}) is provided in Section S3 of the Supplementary Material.  Finally, for a given estimator of $H_0$, denoted by $\widehat{H}_0$, the pseudo maximum likelihood estimator of $(\gamma,\theta)$ is defined to be the arguments which maximize  $L(\gamma,\theta,\widehat{H}_0)$. 

Estimation of $H_0$ will be done by applying the innovation theorem \citep[][Theorem 3.4]{aalen1978nonparametric}. We start with defining three counting processes. Let $\tau$ be the maximal follow-up time, and for any 
$t \in [0,\tau]$ and $i=1,\ldots,n$ define the counting processes
$$
N_{i(12)}(t) = \delta_{1i} I(V_i \leq t) \, , \,\,\,\,
N_{i(13)}(t) = \delta_{2i} I(V_i \leq t) \,\,\,\, 
\mbox{and} 
\,\,\,\, 
N_{i(23)}(t) = \delta_{3i} I(W_i \leq t) \, .
$$ 
A key assumption is that given the covariates and the frailty variate, $N_{i(12)}$ and $N_{i(13)}$ are independent. Each $N_{i(jk)}(t)$ has intensity process 
$$
Y_{i(j)}(t) \alpha_{jk}(t|Z_{i}) \omega_i \, = Y_{i(j)}(t) h_{0jk}(t) \alpha^*_{jk}(t|Z_{i}) \omega_i \,\,\,\,\,  jk=12,13,23 \,\,\,\,  i=1,\dots,n
$$
where $Y_{i(1)}(t) = I(V_i \geq t)$ and $Y_{i(2)}(t)=\delta_{1i} I(V_i \leq t \leq W_i)$.
The 
$\sigma$-algebra generated by the observed history up to time $t$, related to $jk=12,13$, denoted by $\mathcal{F}^{(1)}_{t}$, is defined by
$$
\mathcal{F}^{(1)}_{t} = \sigma\{N_{i(12)}(s),N_{i(13)}(s),Y_{i(1)}(s),Z_i;i=1,\ldots,n;0\leq s\leq t\} \, .
$$
The $\sigma$-algebra related to $jk=23$ consists of those observations that were diagnosed with the disease, namely, those  with $\delta_{1i}=1$ and $\delta_{2i}=0$, so it is defined by
$$
\mathcal{F}^{(2)}_{t} = \sigma\{V_i, N_{i(23)}(s),Y_{i(2)}(s),Z_i;i \in \{1,\ldots,n\} \,\, \mbox{such that} \, \delta_{1i}=1, \delta_{2i}=0 ;0\leq s\leq t\} \, .
$$
By the innovation theorem, the stochastic intensity process of $N_{i(1k)}(t)$, $k=2,3$, with respect to $\mathcal{F}^{(1)}$, is given by
\begin{equation*}\label{eq:stochint1}
Y_{i(1)}(t) \alpha_{1k}(t|Z_{i})  E(\omega_i|\mathcal{F}^{(1)}_{t-}) = Y_{i(1)}(t) h_{01k}(t)\alpha^*_{1k}(t|Z_{i})  E(\omega_i|\mathcal{F}^{(1)}_{t-})\, ,
\end{equation*}
where
\begin{eqnarray*}
E(\omega_i|\mathcal{F}^{(1)}_{t})  &=& \int \omega dF(\omega|\mathcal{F}^{(1)}_{t}) =
\frac{\int \omega^{N_{i(1.)}(t)+1 }
	\exp[ -\omega \{ A_{i(1.)}(t|Z_i) \}  ]dF(\omega)}
{\int \omega^{N_{i(1.)}(t)}
	\exp[ -\omega \{ A_{i(1.)}(t|Z_i) \}  ]dF(\omega)}
\nonumber \\
&=&  \frac{(-1)^{N_{i(1.)}(t)+1}\phi^{N_{i(1.)}(t)+1}
	(A_{i(1.)}(t|Z_i))}
{(-1)^{N_{i(1.)}(t)}\phi^{N_{i(1.)}(t)}(A_{i(1.)}(t|Z_i))} \, ,
\end{eqnarray*}
$N_{i(1.)} (t)= N_{i(12)} (t) + N_{i(13)} (t)$ and $A_{i(1.)} (t|Z_i)= A_{12} (t \wedge V_i|Z_{i}) + A_{13} (t \wedge V_i|Z_{i})$.
For subjects with $\delta_{1i}=1$ and $\delta_{2i}=0$, the stochastic intensity process of $N_{i(23)}(t)$, with respect to $\mathcal{F}^{(2)}$, is given by
\begin{equation*}\label{eq:stochint2}
Y_{i(2)}(t) \alpha_{23}(t|Z_{i})  E(\omega_i|\mathcal{F}^{(2)}_{t-}) = Y_{i(2)}(t) h_{023}(t)\alpha^*_{23}(t|Z_{i})  E(\omega_i|\mathcal{F}^{(2)}_{t-})\, ,
\end{equation*}
where
\begin{eqnarray*}
E(\omega_i|\mathcal{F}^{(2)}_{t})  = \frac{\int 
\omega^{2+N_{i23}(t)}
\exp[ -\omega \{ A_{i.}(t|Z_{i}) \}  ]dF(\omega)}
{\int \omega^{1+N_{i23}(t)}
\exp[ -\omega \{ A_{i.}(t|Z_{i}) \}  ]dF(\omega)}
 = \frac{(-1)^{2+N_{i23}(t)}\phi^{(2+N_{i23}(t))}
(A_{i.}(t|Z_{i}))}
{(-1)^{1+N_{i23}(t)}\phi^{(1+N_{i23}(t))}(A_{i.}(t|Z_{i}))} \, ,
\end{eqnarray*}
and $A_{i.}(t|Z_i)=A_{i(1.)}(V_i|Z_{i})+A_{23}(W_i \wedge t|Z_{i})-A_{23}(V_i|Z_{i}) $.

For example, under the gamma frailty model, 
$$
{E}(\omega_i|\mathcal{F}^{(1)}_{t}) = \frac{\theta^{-1}+N_{i(1.)}(t)}{\theta^{-1}+A_{i(1.)}(t|Z_i)} \, ,
$$
and for subjects with $\delta_{1i}=1$ and $\delta_{2i}=0$,
$$
{E}(\omega_i|\mathcal{F}^{(2)}_{t}) = 
\frac{\theta^{-1}+1+N_{i(23)}(t)}{\theta^{-1}+A_{i.}(t|Z_i)} \, .
$$ 

Then, the respective Breslow-type estimators of $H_{0jk}(\cdot)$, $jk=12,13,23$, are defined as step functions with jumps at the respective  observed failure times. That is, 
\begin{equation}\label{eq:cumhaz}
\widehat{H}_{0jk}(t) = \sum_{s \leq t}  \Delta \widehat{H}_{0jk}(s) \, , \,\,  jk=12,13,23  \, ,
\end{equation}
with
\begin{equation}\label{eq:haz12}
\Delta \widehat{H}_{01k}(t) = \frac{\sum_{i=1}^n \delta_{k-1 \, i}I(V_i=t)}
{\sum_{i=1}^n Y_{i(1)}(t) \widehat{\alpha}_{1k}^*(t-|Z_{i}) \widehat{E}\left(\omega_i|\mathcal{F}^{(1)}_{t-}\right) } \,\,\, k=2,3 ,
\end{equation}
and
\begin{equation}\label{eq:haz23}
\Delta \widehat{H}_{023}(t) = \frac{\sum_{i=1}^n \delta_{3i}I(W_i=t)}
{\sum_{i=1}^n Y_{i(2)}(t) \widehat{\alpha}_{23}^*(t-|Z_i) \widehat{E}\left(\omega_i|\mathcal{F}^{(2)}_{t-}\right) } \, ,
\end{equation}
where in $\widehat{E}(\omega_i|\mathcal{F}^{(j)}_{t-})$, $j=1,2$, and in $\widehat{\alpha}^{*}_{jk}$, $jk=12,13,23$, the unknown parameters are replaced by their estimators. A detailed description of $\widehat{A}_{jk}$, the estimators of ${A}_{jk}$, $jk=12,13,23$,
is provided in Section S4 of the Supplementary Material.

The proposed estimation procedure is summarized as follows.
\begin{description}
\item[Step 1.] Use standard Cox regression software to obtain initial values of $\widehat{\gamma}_{12}$, $\widehat{\gamma}_{13}$ and $\widehat{\gamma}_{23}$, by running three separate models, and take $\widehat{\theta}$ to a value near independence.
\item[Step 2.] Use the current values of $(\widehat{\gamma}^T,\widehat{\theta})$ and estimate $H_{0jk}$,
$jk=12,13,23$, by Eq.~(\ref{eq:cumhaz})--(\ref{eq:haz23}).
\item[Step 3.] Use the current estimate $\widehat{H}_{0jk}$, $jk=12,13,23$, and estimate $(\gamma^T,\theta)$ by maximizing 
$L(\gamma,\theta,\widehat{H}_{0})$.
\item[Step 4.]
Iterate between Steps 2 and 3 until convergence is reached.
\end{description}

Let $\mu = (\gamma^{T},\theta)^T$, $\widehat{\mu} = (\widehat{\gamma}^{T},\widehat{\theta})^T$, $\mu^o = (\gamma_{12}^{oT},\gamma_{13}^{oT},\gamma_{23}^{oT},\theta^o)^T$, and $H^o_0 = (H_{012}^o,H_{013}^o,H_{023}^o)$, where the superscript $o$ denotes the respective true value. The following theorem summarizes the asymptotic properties of the proposed estimators. The required technical conditions and a sketch of the proof are provided in Section S5 of the Supplementary Material. 

\begin{theorem} Under the assumptions listed in Appendix A.2, $\widehat{\mu}$ is a consistent estimator of $\mu$, $\sup_t |\widehat{H}_{0jk}(t)-H^o_{0jk}(t)|=O_p(n^{-1/2})$, $jk=12,13,23$, 
$\sqrt{n} (\widehat{\mu} - \mu^o)$ is asymptotically mean-zero multivariate normal, and $\sqrt{n}\{\widehat{H}_{0jk}(t)-H^o_{0jk}(t)\}$, $jk=12,13,23$, converges to a Gaussian process.
\end{theorem}

\subsection{Variance Estimation}
Deriving the asymptotic or finite-sample variances of the proposed estimators analytically, is challenging, and is not attempted here. Instead, we advocate the use of the weighted bootstrap approach \citep{kosorok2004robust}. Within each bootstrap sample, a random weight is assigned to each observation, from a standard exponential distribution. The estimators of each bootstrap sample are then derived based on $\log L^{(b)}(\gamma,\theta,H_0) = \sum_{i=1}^n\eta^{(b)}_i \log(L_i)$, where $\eta^{(b)}_i$ is the weight for subject $i$ of the $b$-th bootstrap repetition. Likewise, the $b$-th bootstrap estimation of the baseline hazard function $H_{012}(t)$ consists of  
$$\Delta \widehat{H}^{(b)}_{012}(t) 
= \frac{\sum_{i=1}^n \eta^{(b)}_i \delta_{1i}I(V_i=t)}
{\sum_{i=1}^n \eta^{(b)}_i Y_{i(1)}(t) \widehat{\alpha}^{*(b)}_{12}(t-|Z_{i}) 
\widehat{E}^{(b)}\left(\omega_i|\mathcal{F}^{(1)}_{t-}\right) } \, ,
$$ and similarly for $H_{013}(t)$ and $H_{023}(t)$. 
The weighted bootstrap approach is more suitable than regular bootstrap in this case, because in highly censored data, the regular bootstrap could produce samples with a low number of events. 

\subsection{Computational aspects}
We analyze the large-scale UKB dataset. Taking CRC as an example, among the 221,723  men (263,195 women) there were 1,603 (1,189) CRC incident cases, 7,752 (5,015) died during the follow-up time before having CRC, and out of the 2945 (2,186) prevalent and incident CRC observations, 668 (372) died. Thus, 212,368 men and 256,991 women were censored.  Our estimation procedure with such a big sample size, is time consuming. Thus, the following is a simple technique for reducing the sample size with only small efficiency loss, in the spirit of the basic ideas used in case-cohort designs \citep{cai2004sample}. In particular, the  log-likelihood based on (\ref{eq:likelihood}) 
can be written as
\begin{eqnarray*}\label{eq:short}
&& \sum_{i=1}^n \left[ \sum_{j=1,2}
 {\delta_{ji}}\log\{h_{01 \, j+1}(V_i) \alpha^*_{1\, j+1}(V_i|Z_{i}) \} +
{\delta_{3i}} \log \{h_{023}(W_i) \alpha^*_{23}(W_i|Z_{i})\} \right]\nonumber \\
&& + 
\sum_{i=1}^n I(\delta_{.i}>0) \log 
\{(-1)^{\delta_{.i}} \phi^{(\delta_{.i})}(s_i) \} 
+
\sum_{i=1}^n I(\delta_{.i}=0) \log 
 \phi(s_i)  \, ,
\end{eqnarray*}
where $s_i$ is given in (\ref{eq:si}).
In the CRC UKB data, the last sum consists of  more than 200,000 observations, within each sex, while most of the information is provided by the events. Let $n_0 = \sum_{i=1}^n I(\delta_{.i}=0) $. Then, for big datasets with high censoring rates, we recommend taking a random sub-sample of size $\tilde{n}$  among the censored observations (i.e. those with $\delta_{.i}=0 $), denoted by $\mathcal{C}$, and the above log-likelihood function is replaced by  
\begin{eqnarray}\label{eq:short2}
&& \sum_{i=1}^n \left[ \sum_{j=1,2}
{\delta_{ji}}\log\{h_{01 \, j+1}(V_i) \alpha^*_{1\, j+1}(V_i|Z_{i}) \} +
{\delta_{3i}} \log \{h_{023}(W_i) \alpha^*_{23}(W_i|Z_{i})\} \right]\nonumber \\
&& + 
\sum_{i=1}^n I(\delta_{.i}>0)\log 
\{ (-1)^{\delta_{.i}} \phi^{(\delta_{.i})}(s_i) \} +
\frac{n_0}{\tilde{n}}\sum_{i \in \mathcal{C}} I(\delta_{.i}=0) \log 
 \phi(s_i)  \, .
\end{eqnarray} 
Similarly, the denominators of the cumulative baseline hazard estimators of $H_{01k}$, $k=2,3$, are replaced by
\begin{eqnarray}\label{eq:haz12short}
&&\sum_{i=1}^n I(\delta_{.i}>0) Y_{i(1)}(t) \widehat{\alpha}_{1k}^*(t-|Z_{i}) \widehat{E}\left(\omega_i|\mathcal{F}^{(1)}_{t-}\right) \nonumber \\
&& +
 \frac{n_0}{\tilde{n}}
\sum_{i \in \mathcal{C}} I(\delta_{.i}=0)
 Y_{i(1)}(t) \widehat{\alpha}_{1k}^*(t-|Z_{i}) \widehat{E}\left(\omega_i|\mathcal{F}^{(1)}_{t-}\right)  \, .
\end{eqnarray} 
There is no change in the estimator of $H_{023}$ since the sub-sampling step has no effect on the observations involved with this estimator. In summary, the sample consists of all the observations with at least one observed event and a random sub-sample from the censored data, where for each observation of the sub-sample a weight of $n_0/\tilde{n}$ is assigned; the rest are assigned with a weight of 1. \cite{cai2004sample} studied the efficiency loss as a function of the failure and sampling rates for a simple case-cohort design. For example, with a failure rate of 0.01 and a sampling rate $\tilde{n}/n_0=0.1$, the relative efficiency loss is less than 0.05. In the following UKB data analysis, $\tilde{n}=20,000$.

\section{Methods:  right-censored data and delayed entry }
\subsection{Data and assumptions}
In addition to the random variables defined previously, we assume that  subject $i$ is recruited at age $R_i$, $c_L \leq R_i \leq c_U$, $i=1,\ldots,n$, and then followed prospectively  until death or censoring, whichever comes first. In the UKB data, $c_L=40$ and $c_U=69$. Thus, the data consist of $n$ independent observations, each with $(V_i,\delta_{1i},\delta_{2i},W_i, \delta_{3i},Z_i,R_i)$.
Some participants had the disease before recruitment, namely $R_i > V_i$, and these observations are referred to as prevalent, whereas those who develop the disease after being recruited, $R_i \leq V_i$, are referred to as incident observations. Such a design, known also as length bias (or left truncation),  suffers from sampling bias since only those individuals who live long enough are observed. 

As there are no incident cases below the age of $c_L$, one cannot directly estimate from the data any of the hazard functions below that age. In this work we focus on CRC. Since having CRC before age 40 is very rare \citep{ouakrim2015trends}, we assume that the probability of having the disease before age $c_L$ is practically zero. Hence, the estimators of the hazard functions of types $jk=12,23$ are only very slightly biased. Such an assumption should not be adopted for diseases such as breast cancer, where approximately 7\% of diagnoses are before the age of 40 years \citep{anders2009breast}. Thus, our proposed analysis is not directly applicable for such phenotypes and additional adjustment is required. Likewise, it is impossible to directly estimate the hazard functions of death before having the disease, $h_{13}(t|Z)$ and $h^c_{13}(t|Z,\omega)$, based on the observed data.

For an illness-death model with delayed entry, there are three principal statistical methods for inference \citep[][and references therein]{vakulenko2016comparing}: (i) an unconditional approach where $R_i$ is considered as a random variable with a known distribution, and its distribution is included in the likelihood function; (ii) a conditional approach where the value of the recruitment age, $R_i$, is conditioned upon; or (iii) a conditional approach where the  entire observed history up to the recruitment age, $R_i$, is conditioned upon. In practice, multivariate survival data with delayed entry are most often analyzed using approach (iii) \citep{andersen1988multistate,saarela2009joint}. In an illness-death model with approach (iii), prevalent individuals are not considered for estimation prior to their entry time, so they only contribute for estimating the parameters related to transition from  diseased state to death. 

Since we have no knowledge or reasonable assumptions on the recruitment distribution (except for its support), approach (i) is inapplicable. While approach (ii) is more efficient than (iii) since it utilizes more information, it is more challenging computationally, as it requires an additional complicated numerical integration. In this section we  propose an estimation procedure adapted for delayed entry, which is based on the procedure of Section 2, and applies approach (iii). 

Our proposed estimation procedure for accommodating delayed entry consists of three modifications: (1) adjusting the likelihood; (2) leveraging external information to estimate the baseline hazard function of disease-free death, $h_{013}$, at age $t < c_L$; and (3) adjusting the hazard functions estimators, $H_{0jk}$, $jk=12,13,23$. 

\subsection{Adjusted likelihood} 
The likelihood function based on the observed data given the history up to entry age, $R_i$, $i=1,\ldots,n$, is given by
$L^{LT}(\gamma,\theta,H_0) = \prod_{i: R_i<V_i} L_i^{LT1} \prod_{i: R_i>V_i} L_i^{LT2}$, where 
\begin{eqnarray*}
L_i^{LT1} &\propto& 
\{h_{012}(V_i) \alpha^*_{12}(V_i|Z_{i}) \}^{\delta_{1i}}
\{h_{013}(V_i) \alpha^*_{13}(V_i|Z_{i})\}^{\delta_{2i}} 
\{h_{023}(W_i) \alpha^*_{23}(W_i|Z_{i})\}^{\delta_{3i}} \nonumber \\
& & 
(-1)^{\delta_{.i}} \phi^{(\delta_{.i})}(s_{11i})  /\phi(s_{12i}) \, ,
\end{eqnarray*} 
$s_{11i} = A_{12}(V_i|Z_{i})+A_{13}(V_i|Z_{i})+\delta_{1i}A_{23}(W_i|Z_{i}) - \delta_{1i} A_{23}(V_i|Z_{i})$ 
and $s_{12i} = A_{12}(R_i|Z_{i})+A_{13}(R_i|Z_{i})$.
Also,
\begin{eqnarray*}
L_i^{LT2} \propto \{h_{023}(W_i) \alpha^*_{23}(W_i|Z_{i})\}^{\delta_{3i}} 
(-1)^{1+\delta_{3i}} \phi^{(1+\delta_{3i})}(s_{21i})  /(-1) \phi^{(1)}(s_{22i}) \, ,
\end{eqnarray*}
where
$s_{21i} = A_{12}(V_i|Z_{i})+A_{13}(V_i|Z_{i})+A_{23}(W_i|Z_{i}) - A_{23}(V_i|Z_{i})$
and
$s_{22i} = A_{12}(V_i|Z_{i})+A_{13}(V_i|Z_{i})+A_{23}(R_i|Z_{i})- A_{23}(V_i|Z_{i})$. Details on the derivation of the above formulas are provided in Section S6 of the Supplementary Material.

\subsection{Estimating $h_{013}(t)$ for age $t < c_L$}

Estimation of the hazard functions under delayed entry is usually done in one of two approaches: risk-set correction or inverse probability weighting (IPW). In the risk-set correction approach, prevalent cases cannot contribute to the estimation of $h_{12}(t|Z)$, $h^c_{12}(t|Z,\omega)$, $h_{13}(t|Z)$, $h^c_{13}(t|Z,\omega)$, since only observations that satisfy the condition $R_i \leq V_i \leq t$ are included in the risk set; but they can nevertheless contribute to the estimation of $h_{23}(t|Z)$ and $h^c_{23}(t|Z,\omega)$. By contrast, with the IPW approach, prevalent cases can contribute more in some settings. However, in our setting this approach is inapplicable. The IPW approach is based on the idea that observations with a small sampling probability are given more weight so as to rectify their under-representation in the data. Since in the UKB data only observations who die after the age of 40 can be included in the first place, those who died before that age have a sampling probability of 0, and the IPW cannot overcome it. Therefore, we use the risk-set correction approach. 

We propose to estimate $H_{013}$ for $t \leq c_L$ by leveraging the external information on death rate in the general population, for example from life tables. We assumed that the marginal death distribution in the general population approximates sufficiently the marginal death distribution among individuals free of the disease (a reasonable assumption for diseases that are rare among individuals of age $c_L$ or less), and that there exists a reasonable comparability between the general population and the UKB population.
The first step is to use general population data to estimate the marginal hazard $h_{13}(t)$ for $t \leq c_L$. For our analysis, we have used data published by the UK Office for National Statistics (https://www.ons.gov.uk). 

Proceeding further, the marginal survival function of $T_{13}$ can be expressed as
$$
S_{13}(t) = \int \exp\{ -H_{013}(t-)\exp(\gamma_{13}^T z)\} f(z)dz \,
$$
and, differentiating with respect to $t$, the marginal density function is seen to be equal to
$$
f_{13}(t) = S_{13}(t) h_{13} (t) =  h_{013}(t) \int \exp(\gamma_{13}^T z)  \exp\{ -H_{013}(t-)\exp(\gamma_{13}^T z) \} f_Z(z)dz \, . 
$$
Hence, the relationship between $h_{13}(t)$ and 
$h_{013}(t)$ is 
\begin{eqnarray*}
h_{013}(t) = h_{13}(t) \frac{\int \exp\{ -H_{013}(t-)\exp(\gamma_{13}^T z)\}f_Z(z)dz}{\int \exp(\gamma_{13}^T z)  \exp\{ -H_{013}(t-)\exp(\gamma_{13}^T z) \} f_Z(z)dz} \, .
\end{eqnarray*}
Assuming that $Z$ in the cohort is representative of its distribution in the population, then, given $\widehat{\gamma}_{13}$ an estimator for $h_{013}(t)$ can be defined as 
\begin{eqnarray}\label{eq:lifetable}
\widehat h_{013}(t) = h_{13}(t) \frac{\sum_{i=1}^n \exp\{-\widehat H_{013}(t-)\exp(\widehat{\gamma}^T_{13} z_i)\} }{\sum_{i=1}^n \exp(\widehat{\gamma}^T_{13} z_i) \exp\{-\widehat H_{013}(t-)\exp(\widehat{\gamma}^T_{13} z_i)\}}, 
\end{eqnarray}
where $\widehat h_{013}(t)$ is estimated successively from 0 to $c_L$ at pre-specified equally-spaced  grid of $\kappa$ points of $h_{13}(t)$, and $\widehat H_{013}(t)=c_L/\kappa \sum_{u \leq t} \widehat h_{013}(u)$. If recruitment starts at age $c_L$, the estimator of $H_{013}$ will be based on (\ref{eq:lifetable}) up to age $c_L$, and then will continue with the following estimator provided in Section 3.4.  

For diseases where the probability of onset before $c_L$ is not negligible, such as breast cancer, a similar approach can be implemented upon the availability of similar disease incidence information in order to estimate $h_{12}$ and $h_{12}^c$ before $c_L$. 

\subsection{The adjusted hazard function estimators}
For estimating the cumulative hazard functions, the intensity processes above are used while correcting the risk-sets. Specifically, the respective Breslow-type estimators of $H_{0jk}(\cdot)$, $jk=12,13,23$, are 
\begin{equation}\label{eq:hazLT1}
\Delta \widehat{H}^{LT}_{012}(t) = \frac{\sum_{i=1}^n I(R_i<V_i)\delta_{1i}I(V_i=t)}
{\sum_{i=1}^n I(R_i \leq t \leq V_i) \widehat{\alpha}_{12}^{*}(t-|Z_i) \widehat{E}\left(\omega_i|\mathcal{F}^{(1)}_{t-}\right) } \,\,\,\,\, t >0 \, ,
\end{equation}
\begin{equation}\label{eq:hazLT2}
\Delta \widehat{H}^{LT}_{013}(t) = \frac{\sum_{i=1}^n I(R_i<V_i)\delta_{2i}I(V_i=t)}
{\sum_{i=1}^n I(R_i \leq t \leq V_i) \widehat{\alpha}_{13}^{*}(t-|Z_i) \widehat{E}\left(\omega_i|\mathcal{F}^{(1)}_{t-}\right) } \,\,\,\,\, t \geq c_L \, ,
\end{equation}
and for $t>0$,
\begin{equation}\label{eq:hazLT3}
\Delta \widehat{H}^{LT}_{023}(t) = \frac{\sum_{i=1}^n \delta_{3i}I(W_i=t)}
{\sum_{i=1}^n  I(R_i < t) Y_{i(2)}(t) \widehat{\alpha}_{23}^{*}(t-|Z_i) \widehat{E}\left(\omega_i|\mathcal{F}^{(2)}_{t-}\right) } \, .
\end{equation}
\\

To summarize, the following are the updated Steps 2 and 3 of the proposed estimation procedure for delayed-entry and right-censored data (Steps 1 and 4 are the same as before):
\begin{description}
\item[Step $\bf \widetilde 2$.] Use the current values of $(\widehat{\gamma}^T,\widehat{\theta})$ and estimate $H_{0jk}$,
$jk=12,13,23$, by Eq.'s (\ref{eq:lifetable}) -- (\ref{eq:hazLT3}).
\item[Step $\bf \widetilde 3$.] Use the current estimate $\widehat{H}_{0jk}$, $jk=12,13,23$, and estimate $\gamma$ and $\theta$ by maximizing 
$L^{LT}(\gamma,\theta,\widehat{H}_0)$.
\end{description}

In order to estimate the estimators variance, we suggest using the weighted bootstrap, as described in Section 2.2.

\section{Analysis of UKB Colorectal Cancer (CRC) Data}
\subsection{Data Processing}

The failure time related to CRC, defined to be the age at first invasive colorectal cancer diagnosis, and death from colorectal cancer, were according to the ICD10 codes (C180, C182-C189, C19, and C20) or the ICD9 codes (1530--1534, 1536--1541). Cancer of the appendix or non-invasive (in situ) colorectal cancer cases were excluded, as well as cases of carcinoid or related tumors (8240--8249) or lymphomas (9590--9729). 

To protect the participants' anonymity, some information, such as exact birth dates, is suppressed from the dataset. Whenever we could calculate the exact recruitment ages, we did so, typically for observations who were diagnosed with cancer, or that have died. We were able to calculate those ages since exact dates of cancer diagnosis and death are provided in the data, in addition to the exact recruitment dates. Whenever we were unable to procure the exact recruitment ages, we arbitrarily set the birth dates of those observations to the 15-th of the month they were reported to be born in. 

\subsection{Anaylsis Results}
We followed the analysis of \citet{jeon2018determining}, and generated an environmental risk score (E-score), for lifestyle and environmental risk factors. Specifically, a Cox model with a delayed-entry adjustment was fitted, with the age at diagnosis of CRC as the outcome and the recruitment age was used for the risk-set correction. The following well-known CRC risk factors were included as covariates: sex, height, body mass index, education, smoking status, alcohol consumption, ibuprofen use, drugs use, use of post-menopausal hormones (women only), and physical activity. Prevalent CRC cases were excluded. The results are provided in Table S1 of the Supplementary Material. 

The E-score of each participant, is then defined as a linear combination of all risk factors, each one weighted by its estimated regression coefficient. 
The E-scores were subsequently standardized by performing a quantile transformation based on the E-score empirical cumulative distribution function of the CRC-free observations. The transformed E-scores were then entered into the following illness-death model as a covariate. The bottom of Table S1 provides the means and standard deviations of the transformed E-scores, by CRC status and sex.

Similarly, a genetic risk score (G-score) was derived, based on 72 single-nucleotide polymorphisms (SNPs) that have been identified  to be associated with CRC by GWAS \citep{jeon2018determining}. Each SNP variable was coded as dosage, which is 0,1, or 2, based on the number of risk allele copies if the SNP is directly genotyped, and expected number of copies if it is imputed. The G-score was developed in a similar manner to the E-score. Specifically, a similar Cox regression model was fitted on the CRC onset age as the outcome, and the 72 SNPs as covariates. The G-score was constructed for each subject as the weighted sum of the 72 SNPs, with the estimated regression coefficients as weights. A quantile transformation was once again used, based on the G-score empirical cumulative distribution function of the CRC-free observations. The transformed G-scores were then entered into our proposed illness-death model as a covariate. A detailed description of the SNPs and the analysis results
are provided in Tables S2--S3 of the Supplementary Material.

Women have a much lower risk of CRC. To allow more precise estimates of the baseline hazard functions, our proposed illness-death model with delayed-entry adjusted estimation procedure was applied separately to men and women. The regression coefficient vectors $\gamma_{12}$  and $\gamma_{13}$ included G-score and E-score, while $\gamma_{23}$ included just the G-score. The E-score is not included in the transition model $2 \rightarrow 3$ since this part of the model was estimated using both prevalent and incident data, and environmental data on the prevalent cases were expected to be subject to substantial recall bias. 
We compared our method with the following Cox models (in which the disease onset age was, in the spirit of Shen, 2017, included in the model for the transition $2 \rightarrow 3$):
\begin{itemize}
\item[Cox I:] Three separate Cox models were fitted. In particular, $\gamma_{12}$ and $H_{012}$ are estimated based on CRC age at diagnosis as the outcome, age at recruitment is used for risk-set correction, and other events are treated as independent censoring; $\gamma_{13}$ and $H_{013}$ are estimated based on age at death before CRC as the outcome, age at recruitment is used for risk-set correction, other events are treated as independent censoring; $\gamma_{23}$ and $H_{023}$ are estimated based on age at death after CRC as the outcome, and age at CRC diagnosis and age at recruitment are used for risk-set correction. See Section S8 of the Supplementary Material, for the partial likelihoods.
\item[Cox II:]  $\{\gamma_{12}, H_{012}, \gamma_{13}, H_{013}\}$ are estimated as in Cox I. In $\gamma_{23}$ the standardized age at diagnosis is added as a time-independent covariate, for dealing with the fact that $V$ is a dependent left-truncation time \citep{shen2017semiparametric}.
\item[Cox III:]  $\{\gamma_{12}, H_{012}, \gamma_{13}, H_{013}\}$ are estimated as in Cox I. The effect of age of CRC diagnosis is included using a linear truncated spline with three knots, at the 25\%, 50\%, and 75\%. 
\end{itemize}

\begin{figure}[!t]
	\begin{center}
		\includegraphics[width=0.8\linewidth]{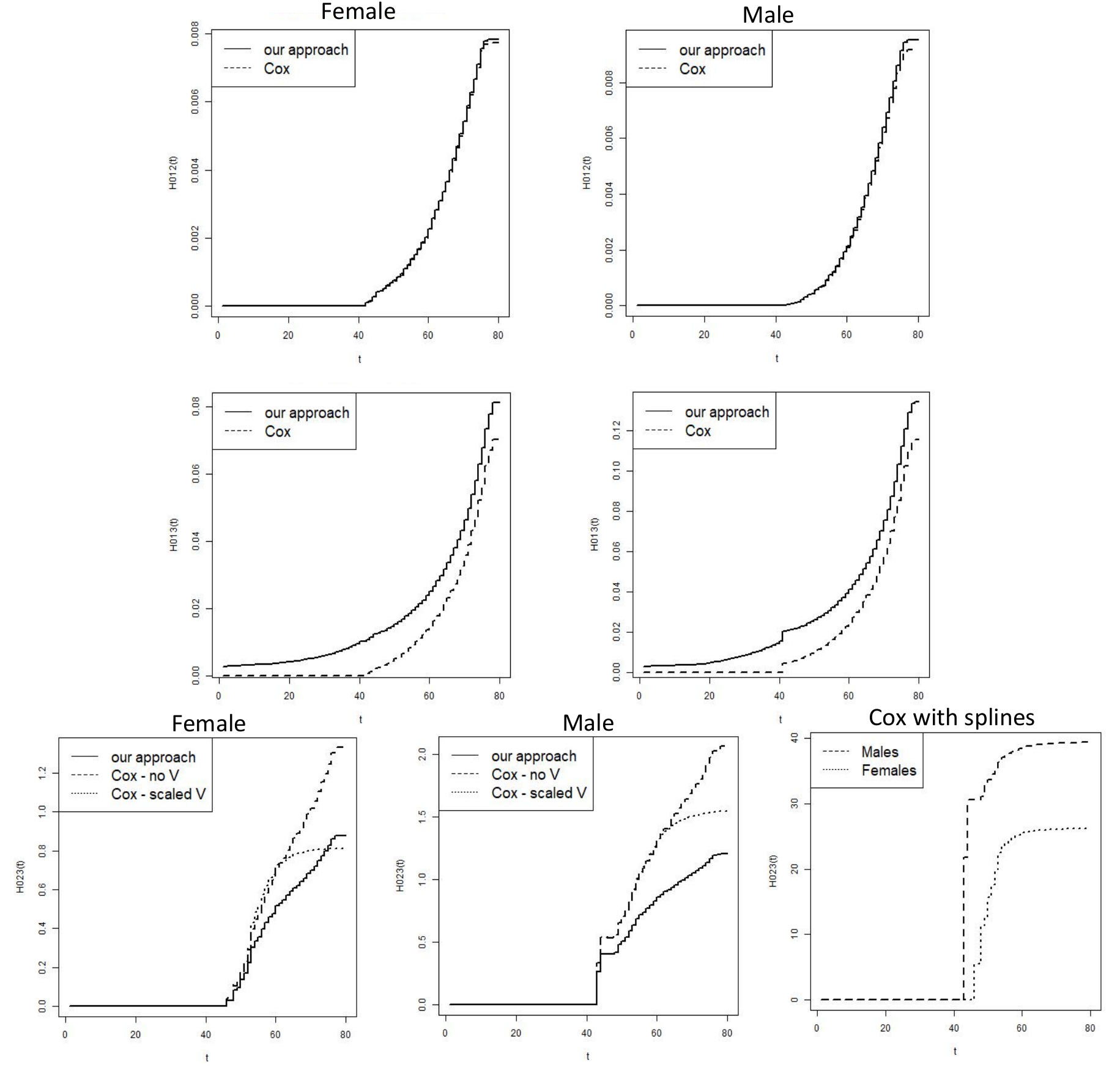}
		\caption{Baseline hazard functions of UKB colorectal cancer data. First two rows are the estimators of $H_{012}$ and $H_{013}$, respectively, women on the left and men on the right.  The third row is the results of $H_{023}$.  \label{hazfig}}
	\end{center}
\end{figure}

The results are presented in Table~\ref{restable} and Figure~\ref{hazfig}. As expected, the Cox and the proposed estimators of $\{\gamma_{12}, H_{012}, \gamma_{13}\}$,  are similar. Under Cox, the baseline hazard function $H_{013}$ equals 0 for $t \leq 40$, thus the Cox estimator of $H_{013}$ is smaller than the proposed estimator. There are substantial differences among the estimators of $H_{023}$, with extreme results under Cox with linear truncated spline. The G-score and E-score coefficients are both significantly greater than zero for the healthy-diseased process. Additionally, it turns out that the G-score for CRC does not bear a significant effect on the healthy-death process, but the corresponding E-score does. This result seems plausible because many CRC-related risk factors such as smoking status and alcohol consumption, are known to be related to death in general, and not only to CRC. Also, it is well-known that women have lower risk for CRC, which might explain why the regression coefficient of the E-score for women in the healthy-diseased model is much smaller than that of men. 

Under the Cox models, the effect of the G-score of the diseased-dead process is null for men, and negative for women. Namely, among the women, based on the Cox analysis, the G-score increases CRC risk, but decreases mortality after having the disease. This result is counter intuitive. In contrast, our proposed analysis suggests that the G-score also tends to increase the risk of death after having the disease. The dependence parameter among the event times is approximately 2 (corresponding to a Kendall tau value of approximately 0.5), in both men and women. The large standard error of $\widehat{\theta}$ among men (0.480) was driven by few outliers in the bootstrap sample, while the median absolute deviation was 0.204. As a result we can deduce that there is a non-negligible dependence between the processes which was not accounted for through the covariates. Moreover, under the Cox model, the assumption of independent left-truncated time for $h_{23}$ is most likely violated and thus yields biased estimates.

\section{Simulation Study}
\subsection{Data Generation}
An extensive simulation study was performed to demonstrate the finite-sample properties of the proposed estimation procedures, with and without delayed entry.

We assumed a gamma frailty model with expectation 1 and variance $\theta$, and a covariate vector $Z^T = (Z_1,Z_2,Z_3,Z_4)$, each covariate generated independently from $\mbox{Uniform}(0,1)$, $c_L = 0.05$ and $c_U = 0.15$. Given a sample of $n$ frailty values and covariate vectors, the failure times $T_{1}$ and $T_{2}$ were generated based on the conditional models $h^c_{jk}$, $jk=12,13,23$. The marginalized baseline hazard functions were set to be
\begin{align*}
h_{012}(t)&=0.005I(0\leq t < 0.05) + I(t \geq 0.05) \\
h_{013}(t)&=0.5I(0\leq t < 0.05) + I(0.05 \leq t \leq 0.15) + 2I(t \geq 0.15) \, \\ 
\text{and, } h_{023}(t) &=  I(t \geq 0.12).
\end{align*}
The regression coefficients were chosen to be $\gamma^T_{12} = (2,0.2,0.05,0),\gamma^T_{13} = (0.05,1,0,0)$ and $\gamma^T_{23}=(1,0,0,0.5)$. Three levels of dependence were studied, $\theta=0, 1$ or 2, corresponding to Kendall tau values of approximately 0, 0.35 and 0.5, respectively. 
The motivation behind the independence configuration, $\theta=0$, is for exploring whether using our procedure results in a substantial efficiency loss compared to the standard partial likelihood approach corrected for delayed entry.

Using the  values of $\alpha_{jk}, jk=12,13$, $T_1$ and $T_2$ were generated by solving for $T$ the equation
$
\exp \{ - A_{jk}(T | Z) \omega \} = U
$,
where $U$ is a value generated from $\mbox{Uniform}(0,1)$. For individuals that were diagnosed with the disease, the original $T_2$ values were discarded, and given $T_1$, $Z$ and $\omega$, a new $T_2$ value was sampled from the respective truncated distribution, by solving for $T$ the equation
$
\exp \{ - A_{23}(T | Z) \omega \} = U\exp \{ - A_{23}(T_1 | Z) \omega \}.
$
Without delayed entry, all observations were followed since time 0. 
Under delayed entry settings, recruitment ages $R$ were randomly generated from  $\mbox{Uniform}(c_L,c_U)$. For each sample, a large dataset consisting of $\left\{T_{1},T_{2},R \right\}$ was first generated, and from those who were still alive at their recruitment age $R$, $n$ observations were randomly sampled. Next, for each observation an independent censoring time was generated from an exponential distribution with a rate parameter of 2. In addition, an administrative censoring was imposed at time 0.61. For $\theta=1$ about 25\% of observations are censored before disease onset or death, and among those who were diseased, about 65\% are censored before death. For $\theta=2$ the corresponding numbers are about $27\%$ and $75\%$. 

%For exploring the effect of model misspecification, we generated data with frailty effects coming from an Inverse Gaussian (IG) distribution and analyzed the data as if the frailty distribution were Gamma. The density function of the IG distribution with mean 1 is $f(\omega)=\sqrt{\frac{\theta}{2\pi \omega^3}}\exp[\frac{-\theta (\omega-1)^2}{2\omega}]$, $\omega>0$. The value of the IG parameter was set to provide the same Kendall $\tau$ values as those created by the Gamma distribution. Specifically, the respective IG parameter values for $\theta=1,2$ under Gamma, are 0.4 and 0.01 respectively, under IG. 

\subsection{Simulation Results}
The following results are based on 100 repetitions for each configuration, and a sample size of 5000 individuals. The pseudo-likelihood function was maximized with the L-BFGS-B algorithm, as implemented in the optim function in R. Convergence of the algorithm was reached once the relative change in the pseudo-log-likelihood between two consecutive iterations went below 0.0001. 

Under the cohort setting with no delayed entry (see Tables S4--S5 of the Supplementary Material), the proposed and the conventional Cox model yielded unbiased estimators for the parameters of $h_{12}$ and $h_{13}$; however, the conventional Cox model analysis yields biased parameter estimates in $h_{23}$, whereas the proposed estimator is unbiased. For example, with $\theta=2$,  the corresponding estimates of $\gamma_{23,1}=1$ by Cox and the proposed approach are 0.628 (SE=0.172) and 1.037 (SE=0.181). Also, the true values of $H_{023}(t)$ at $t=0.2,0.4,0.6$, are 0.08, 0.28 and 0.48, while the corresponding estimates of Cox are 0.056, 0.164 and 0.257 (SEs are 0.010, 0.026 and 0.039). The respective numbers based on the proposed approach are  0.080, 0.280 and 0.484 (SEs are 0.013, 0.042 and 0.072). Evidently, the proposed approach performs well in terms of bias, and under $\theta=0$ the efficiency loss is minimal, if any. These results highlight the importance of the independent assumption of left truncation, which is violated in the conventional Cox analysis. 

Tables 2--3 summarize the empirical means and standard deviations of $\widehat{\theta}$, the regression coefficients, and the baseline hazard functions at certain time points, under delayed entry. We contrast between Cox with delayed-entry adjustment by the risk-set approach, when relevant (see Section S8 of the Supplementary Material for more details) and the proposed methods. Biased results are displayed in bold. 

Table 2 presents  the biased Cox-model estimator of $\gamma_{23,1}$ under $\theta>0$. The estimator of $\gamma_{23,4}$ is practically unbiased as its corresponding covariate is independent of all the other covariates in the model, while the corresponding covariate of $\gamma_{23,1}$ is also affecting the disease age at onset. Table 3 presents the biased Cox-model results of $H_{013}$ under any value of $\theta$, due to also ignoring the fact that no death can be observed before $c_L=0.05$. 
Cox estimators of $\gamma_{12}$ and $\gamma_{13}$ are unbiased, as expected.
In contrast,  our proposed estimators perform well in terms of bias and coverage rates. In addition, we observe that applying our method under the independence scenario does not result in any substantial efficiency loss. 

%Under frailty misspecification, one can notice that as long as the dependence is mild, our estimators are rather robust and still perform relatively well. As the dependence becomes stronger, the performance starts to deteriorate, as is reflected both in the parameter estimates of table \ref{sim_tta_0}--\ref{sim_tta_2}, and in the baseline hazard functions estimates in figure \ref{cumhazplotsIG}. ...

\section{Discussion}
We proposed a novel semi-parametric, shared-frailty based method for analysing time-to-event data, within the illness-death model framework. Our model accounts for delayed entry and possible dependence between the stochastic processes, and allows for covariates inclusion. The simulation study shows that the procedure works well in terms of bias and variance, and does not suffer from a substantial efficiency loss under the independence scenario. 

An alternative estimation procedure, when there is no delayed entry, could be to estimate $\{\gamma_{12},\gamma_{13},H_{012},H_{013}\}$ by a standard partial likelihood approach, estimating $H_{023}$ based on Eq. (\ref{eq:haz23}), and then plugging these estimates in the likelihood, 
Eq.~(\ref{eq:likelihood}), for estimating $(\theta,\gamma_{23})$, which is a different pseudo likelihood approach. An iterative procedure is still required, between the estimates of   $(\theta,\gamma_{23})$ and that of $H_{023}$. Such a procedure might save some computation time but at the price of some efficiency loss. Under delayed entry, the estimator of $H_{013}$ would require a correction in the spirit of Section 3.2. In a future work, one can compare this approach with our proposed method. Consistency  and asymptotic normality should be worked out from scratch. 

There are a number of directions that this work can be further expanded to. The first direction is to try and develop a method that uses the prevalent cases for estimating $\gamma_{12}$ and $H_{012}$ as well. In addition, instead of using the same frailty effect for all the three processes, a more flexible dependence structure can be considered.   

%Our model can be applied on other datasets as well. Specifically, there are biobanks with much younger ages at recruitment, such as the Vanderbilt's biorepository (BioVU) which uses $c_L=18$. Thus, our proposed method can be applied for analysing such data and phenotypes with early age-at-onset, such as breast cancer, which we did not consider in the context of the UKB.

%\bibliographystyle{biometrika}
\bibliography{literature}

%\newpage

\begin{table}
	\centering
	\spacingset{1.5}
	\footnotesize
	\caption{UKB Analysis Results: regression coefficient estimates (standard error)}
	\begin{tabular}{lcccc}
		& Cox I & Cox II & Cox III & Proposed (20K censored)\\
		\hline	
		\multicolumn{5}{c}{221,723 men; 1,603 CRC incident events; 7,752 deaths before CRC;} \\ 
		\multicolumn{5}{c}{out of the 2,945 with CRC (prevalent and incident) 668 died (out of them, 462 are incident cases) } \\
		$\theta$  &- &    -  & - & 1.957 (0.480) \\
		G-score 12 & 1.358 (0.091)  & 1.358 (0.091) & 1.358 (0.091) & 1.333 (0.087)\\
		E-score 12 & 0.743 (0.089) & 0.743 (0.089) & 0.743 (0.089) & 0.708 (0.095) \\
		G-score 13 & 0.051 (0.039) & 0.051 (0.039) & 0.051 (0.039) & 0.048 (0.050)\\
		E-score 13 & 0.785 (0.041) & 0.785 (0.041) & 0.785 (0.041) & 0.749 (0.050) \\
		G-score 23 & -0.003 (0.139) & -0.072 (0.140) & -0.072 (0.139) & 0.421 (0.131) \\
		Scaled $V$ & - & 1.439 (0.097) & 1.423 (0.219)  & -\\
		Spline $Q_1$ & - & - &  -0.251 (0.414) &  -  \\
		Spline $Q_2$ & - & - &  0.429 (1.536) &  -\\
		Spline $Q_3$ & - & - &  0.004 (0.517) &  -  \\
		\multicolumn{5}{c}{263,195 women; 1,189 CRC incident events; 5,015 deaths before CRC;} \\ 
		\multicolumn{5}{c}{out of the 2,186 with CRC (prevalent and incident) 372 died (out of them, 291 are incident cases) } \\
		$\theta$ & - &    -   & -  & 2.297 (0.161) \\
		G-score 12 & 1.416 (0.106) & 1.416 (0.106) & 1.416 (0.106) & 1.404 (0.143) \\
		E-score 12 & 0.260 (0.101) & 0.260 (0.101) & 0.260 (0.101) & 0.248 (0.105) \\
		G-score 13 & -0.002 (0.049)  & -0.002 (0.049) & -0.002 (0.049) & -0.008 (0.061)\\
		E-score 13 & 0.650 (0.050) & 0.650 (0.050) & 0.650 (0.050) & 0.632 (0.058) \\
		G-score 23 & -0.273 (0.184) & -0.399 (0.184) & -0.396 (0.184)  & 0.208 (0.163) \\
		Scaled $V$  &- & 2.065 (0.158) & 2.424 (0.466) & -\\
		Spline $Q_1$ & - & - &  -0.594 (0.701) &  - \\
		Spline $Q_2$ & - & - &  0.001 (0.707) &  -\\
		Spline $Q_3$ & - & - &  0.535 (0.682) &  -\\
		
		\hline
	\end{tabular}
	\label{restable}
\end{table}

\newpage

\begin{table}[]
	\centering
	\spacingset{1.25}
	\footnotesize 
	\caption{Simulation results of regression coefficients and dependence parameter for right-censored data with delayed entry: Cox is naively corrected for left truncation by $\max(V,R)$ for $jk=23$, and properly corrected for left truncation by $R$ for $jk=12,13$.
	the true parameters values are $\gamma_{12}=(\gamma_{12,1},\gamma_{12,2},\gamma_{12,3},\gamma_{12,4})=(2,0.2,0.05,0)$,
	$\gamma_{13}=(\gamma_{13,1},\gamma_{13,2},\gamma_{13,3},\gamma_{13,4})=(0.05,1,0,0)$, $\gamma_{23}=(\gamma_{23,1},\gamma_{23,2},\gamma_{23,3},\gamma_{23,4})=(1,0,0,0.5)$, $n=5000$. \label{sim_LT}}
	\begin{tabular}{clcccccccc}
	$\theta$  &  & $\theta$ & $\gamma_{12,1}$ &	
	$\gamma_{12,2}$ & $\gamma_{12,3}$ &
	$\gamma_{13,1}$ &	
	$\gamma_{13,2}$ & $\gamma_{23,1}$ & $\gamma_{23,4}$ \\
	\hline 
\multicolumn{9}{c}{Cox Corrected for LT}\\ 
0 & mean & -  & 2.007 & 0.200 & 0.042 & 0.058 & 0.991 & 1.004 & 0.505  \\
& empirical SD & - & 0.103 & 0.084 & 0.078 & 0.089 & 0.093 & 0.110 & 0.105  \\
& bootstrap SE & -  & 0.093 & 0.088 & 0.087 & 0.097 & 0.092 & 0.120 & 0.107  \\
& coverage rate & - & 0.950 & 0.960 & 0.980 & 0.970 & 0.940 & 0.960 & 0.960 \\ 
\multicolumn{9}{c}{The Proposed Approach}\\ 
0 & mean &  0.037 & 1.995 & 0.200 & 0.040 & 0.056 & 0.993 & 1.016 & 0.491\\ 
& empirical SD &  0.058 & 0.090 & 0.091 & 0.093 & 0.094 & 0.091 & 0.126 & 0.109 \\ 
& bootstrap SD & 0.057 & 0.095 & 0.088 & 0.088 & 0.098 & 0.093 & 0.122 & 0.105  \\ 
& coverage rate & 0.990 & 0.970 & 0.930 & 0.940 & 0.970 & 0.960 & 0.950 & 0.920 \\ 
\multicolumn{9}{c}{Cox Corrected for LT}\\ 
1 &  mean & - &  2.000 & 0.198 & 0.046 & 0.045 & 0.993 & \textbf{0.696} & 0.537  \\ 
&  empirical SD & -   & 0.093 & 0.097 & 0.089 & 0.101 & 0.097 & 0.140 & 0.116\\ 
&  estimated SD & - & 0.092 & 0.088 & 0.086 & 0.097 & 0.092 & 0.139 & 0.127   \\
&  coverage rate & - &  0.970 & 0.920 & 0.930 & 0.920 & 0.950 & \textbf{0.370} & 0.980\\ 
	 	     		\multicolumn{9}{c}{The Proposed Approach}\\ 
1  & mean &  1.054 & 2.005 & 0.187 & 0.034 & 0.041 & 0.016 & 1.026 & 0.524\\ 
& empirical SD &  0.139 & 0.095 & 0.093 & 0.075 & 0.128 & 0.111 & 0.115 & 0.111\\ 
& bootstrap SE  & 0.123 & 0.090 & 0.080 & 0.079 & 0.118 & 0.105 & 0.123 & 0.108\\ 
& coverage rate & 0.960 & 0.970 & 0.910 & 0.950 & 0.970 & 0.970 & 0.970 & 0.940     \\
\multicolumn{9}{c}{Cox Corrected for LT}\\ 
2&  mean & -&  1.997 & 0.214 & 0.045 & 0.055 & 1.002 & \textbf{0.585} & 0.520  \\  
& empirical SD & -&   0.105 & 0.093 & 0.084 & 0.099 & 0.099 & 0.162 & 0.151 \\
& bootstrap SD &  - & 0.093 & 0.088 & 0.087 & 0.097 & 0.092 & 0.160 & 0.146 \\ 
& coverage rate &  -&   0.910 & 0.950 & 0.960 & 0.970 & 0.940 &\textbf{ 0.280} & 0.920\\ 
 		\multicolumn{9}{c}{The Proposed Approach}\\ 
2& mean & 2.059 & 1.994 & 0.203 & 0.040 & 0.052 & 0.993 & 1.020 & 0.519 \\ 
& empirical SD &  0.165 & 0.097 & 0.089 & 0.075 & 0.098 & 0.093 & 0.171 & 0.133    \\ 
& estimated SD &  0.170 & 0.096 & 0.083 & 0.080 & 0.097 & 0.088 & 0.153 & 0.128\\ 
& coverage rate &   0.950 & 0.950 & 0.930 & 0.970 & 0.960 & 0.950 & 0.910 & 0.920  \\ 
    \hline
	\end{tabular}
\end{table}

\begin{table}[]
	\centering
	\spacingset{1}
	\footnotesize 
	\caption{Simulation results of the baseline hazard for right-censored data with delayed entry: Cox is naively corrected for left truncation by $V$ for $jk=23$, and properly corrected for left truncation by $R$ for $jk=12,13$, baseline hazard functions at 
	$t=0.1,0.2,0.3,0.4,0.5,0.6$, $n=5000$.
	 The proposed estimation approach and the naive Cox model that ignores the frailty effect but corrects for delayed entry.
	 ESD - empirical SD, CR - empirical coverage rate. \label{sim_LT_Haz}}
	\begin{tabular}{clcccccccccccc}
	$\theta$  &   & 0.1 & 0.2 & 0.3 & 0.4 &	 0.5 & 0.6 & 0.1 & 0.2 & 0.3 & 0.4 & 0.5 & 0.6\\
	\hline 
 & &	\multicolumn{6}{c}{Cox corrected for LT} & \multicolumn{6}{c}{The Proposed Approach}\\
  0 &    $H_{012}(t)$ & 0.050 & 0.150 & 0.250 & 0.350 & 0.450 & 0.550 & 0.050 & 0.150 & 0.250 & 0.350 & 0.450 & 0.550 \\ 
	& mean  & 0.051 & 0.151 & 0.250 & 0.350 & 0.451 & 0.549 & 0.050 & 0.151 & 0.252 & 0.353 & 0.455 & 0.559\\
& ESD  &  0.007 & 0.013 & 0.019 & 0.027 & 0.038 & 0.046 & 0.006 & 0.014 & 0.023 & 0.030 & 0.037 & 0.049\\  
& CR & 0.980 & 0.990 & 0.990 & 0.960 & 0.950 & 0.940 & 0.940 & 0.950 & 0.930 & 0.960 & 0.970 & 0.930   \\
 &  $H_{013}(t)$ & 0.075 & 0.225 & 0.425 & 0.625 & 0.825 & 1.025 & 0.075 & 0.225 & 0.425 & 0.625 & 0.825 & 1.025 \\
	& mean &\textbf{ 0.051} & \textbf{0.201} & \textbf{0.401} & \textbf{0.602} & \textbf{0.803} & \textbf{1.003} & 0.078 & 0.228 & 0.429 & 0.630 & 0.828 & 1.027\\
& ESD  &  0.010 & 0.016 & 0.029 & 0.044 & 0.056 & 0.078 & 0.010 & 0.021 & 0.037 & 0.052 & 0.069 & 0.091\\  
& CR &   \textbf{0.270} & \textbf{0.680} & \textbf{0.830} & \textbf{0.890} & \textbf{0.910} & \textbf{0.920} & 0.950 & 0.960 & 0.920 & 0.910 & 0.930 & 0.940\\
 & $H_{023}(t)$ &  0.000 & 0.080 & 0.180 & 0.280 & 0.380 & 0.480 &  0.000 & 0.080 & 0.180 & 0.280 & 0.380 & 0.480 \\ 
	& mean  & - & 0.081 & 0.181 & 0.281 & 0.381 & 0.482 & - & 0.081 & 0.183 & 0.285 & 0.389 & 0.490\\
& ESD  & - & 0.010 & 0.020 & 0.028 & 0.037 & 0.047 & - & 0.010 & 0.021 & 0.032 & 0.043 & 0.056 \\  
& CR & -  & 0.900 & 0.930 & 0.940 & 0.930 & 0.940 & - & 0.960 & 0.920 & 0.910 & 0.900 & 0.890\\
 1 &    $H_{012}(t)$ &   0.050 & 0.150 & 0.250 & 0.350 & 0.450 & 0.550  & 0.050 & 0.150 & 0.250 & 0.350 & 0.450 & 0.550 \\ 
& mean  & 0.051 & 0.151 & 0.253 & 0.353 & 0.452 & 0.553 &0.050 & 0.152 & 0.252 & 0.354 & 0.455 & 0.560 \\
& ESD   & 0.009 & 0.017 & 0.025 & 0.034 & 0.042 & 0.052 &  0.007 & 0.018 & 0.027 & 0.038 & 0.047 & 0.059 \\  
& CR & 0.930 & 0.900 & 0.930 & 0.930 & 0.930 & 0.950 &  0.930 & 0.960 & 0.950 & 0.950 & 0.960 & 0.960\\
&  $H_{013}(t)$  & 0.075 & 0.225 & 0.425 & 0.625 & 0.825 & 1.025 & 0.075 & 0.225 & 0.425 & 0.625 & 0.825 & 1.025 \\
& mean  & \textbf{0.050} & \textbf{0.203} & \textbf{0.404} & \textbf{0.606} & \textbf{0.807} & \textbf{1.009} &  0.080 & 0.233 & 0.435 & 0.639 & 0.844 & 1.040 \\
& ESD  & 0.009 & 0.018 & 0.031 & 0.045 & 0.061 & 0.079 & 0.011 & 0.022 & 0.039 & 0.059 & 0.082 & 0.101\\  
& CR &   \textbf{0.220} & \textbf{0.660} & \textbf{0.870} & \textbf{0.900} & \textbf{0.910} & \textbf{0.920} &  0.950 & 0.960 & 0.950 & 0.950 & 0.920 & 0.920 \\
& $H_{023}(t)$ & 0.000 & 0.080 & 0.180 & 0.280 & 0.380 & 0.480 &  0.000 & 0.080 & 0.180 & 0.280 & 0.380 & 0.480 \\ 
& mean  & - & \textbf{0.071} & \textbf{0.147} & \textbf{0.218} & \textbf{0.284} & \textbf{0.350} &  - & 0.079 & 0.178 & 0.277 & 0.377 & 0.477\\
& ESD   & - & 0.011 & 0.021 & 0.030 & 0.037 & 0.044 & - & 0.009 & 0.017 & 0.026 & 0.036 & 0.045 \\  
& CR & - & \textbf{0.740} & \textbf{0.550} & \textbf{0.390} & \textbf{0.240} & \textbf{0.180} &-  & 0.940 & 0.930 & 0.980 & 0.980 & 0.980\\
2 &    $H_{012}(t)$ &   0.050 & 0.150 & 0.250 & 0.350 & 0.450 & 0.550 & 0.050 & 0.150 & 0.250 & 0.350 & 0.450 & 0.550 \\ 
	& mean  & 0.050 & 0.150 & 0.250 & 0.350 & 0.449 & 0.549 & 0.051 & 0.152 & 0.252 & 0.353 & 0.454 & 0.556\\
& ESD  &  0.007 & 0.014 & 0.022 & 0.031 & 0.039 & 0.047 & 0.007 & 0.013 & 0.021 & 0.028 & 0.036 & 0.045\\  
& CR &  0.940 & 0.960 & 0.930 & 0.930 & 0.930 & 0.960  & 0.970 & 0.960 & 0.950 & 0.960 & 0.950 & 0.970\\
 &  $H_{013}(t)$ &0.075 & 0.225 & 0.425 & 0.625 & 0.825 & 1.025  & 0.075 & 0.225 & 0.425 & 0.625 & 0.825 & 1.025   \\ 
	& mean  &   \textbf{0.049} & \textbf{0.200} & \textbf{0.400} & \textbf{0.601} & \textbf{0.801} & \textbf{0.999}&  0.077 & 0.229 & 0.431 & 0.633 & 0.836 & 1.038\\
& ESD  & 0.008 & 0.018 & 0.033 & 0.045 & 0.060 & 0.074 & 0.009 & 0.019 & 0.033 & 0.046 & 0.060 & 0.072 \\  
& CR &    \textbf{0.130} & \textbf{0.590} & \textbf{0.800} & \textbf{0.890} & \textbf{0.940} & \textbf{0.940} & 0.980 & 0.980 & 0.930 & 0.960 & 0.950 & 0.960 \\
 & $H_{023}(t)$ &  0.000 & 0.080 & 0.180 & 0.280 & 0.380 & 0.480 & 0.000 & 0.080 & 0.180 & 0.280 & 0.380 & 0.480 \\ 
	& mean  &  - &\textbf{ 0.058} & \textbf{0.116} & \textbf{0.168} & \textbf{0.217} & \textbf{0.264} & - & 0.081 & 0.181 & 0.281 & 0.380 & 0.480\\
& ESD  & - & 0.010 & 0.017 & 0.023 & 0.030 & 0.034  & - & 0.013 & 0.027 & 0.039 & 0.052 & 0.063\\  
& CR &  - & \textbf{0.330} & \textbf{0.090} & \textbf{0.020} & \textbf{0.000} & \textbf{0.000} & - & 0.920 & 0.920 & 0.940 & 0.950 & 0.940\\

    \hline
	\end{tabular}
\end{table}

\newpage
\renewcommand{\thetable}{S\arabic{table}}
\setcounter{section}{0}

\begin{center}
	{\LARGE \textbf{Supplementary Material}}
\end{center}

\spacingset{1.45} % DON'T change the spacing!
\section{Proof of Lemma 1}
The following is the proof of Lemma 1. Let $A_{jk}(t|Z)=\int_0^t \alpha_{jk}(s|Z) ds$, $jk=12,13,23$. Then, for $t>0$, the relationship between $\alpha_{jk}$ and $h_{jk}$, can be derived as follows. Denote by $j \rightarrow k$ the event of transition from state $j$ to state $k$. Then,  for $k=2,3$,
\begin{eqnarray}\label{eq:tran1k}
\Pr( 1 \rightarrow k \in [t,t+\Delta]|Z) &=& \int \Pr( 1 \rightarrow k \in [t,t+\Delta]|Z, \omega) dF(\omega) \nonumber \\
&\doteq & \Delta \int \omega \alpha_{1k}(t|Z) \exp\{-\omega A_{1.}(t|Z) \} dF(\omega) \nonumber \\
&=& -\Delta \phi^{(1)}\{ A_{1.}(t|Z)\} \alpha_{1k}(t|Z)  \, ,
\end{eqnarray}
where $A_{1.}(t|Z) = A_{12}(t|Z)+A_{13}(t|Z)$, $\phi(s)$ is the Laplace transform of $\omega$, and
$\phi^{(q)}(s)$ is the $q$th derivative with respect to $s$.  In addition,
\begin{eqnarray}\label{eq:stay1}
\Pr(\mbox{remain in 1 during} \, [0,t]|Z ) = \int \exp\{ -\omega A_{1.}(t|Z)\} dF(\omega) 
= \phi\{A_{1.}(t|Z)\} 
\end{eqnarray}
At the same time,
\begin{eqnarray}\label{eq:stay2}
\Pr(\mbox{remain in 1 during} \, [0,t]|Z ) = \exp\{-H_{1.}(t|Z)\} \, ,
\end{eqnarray}
where $H_{1.}(t|Z) = H_{12}(t|Z)+H_{13}(t|Z)$ and $H_{jk}(t|Z) = \int_0^t h_{jk}(u|Z) du$.
Then, based on the ratio of (\ref{eq:tran1k}) and (\ref{eq:stay1}) we get the following equations,
\begin{equation}\label{eq:h012A}
h_{12}(t|Z) =  h_{012}(t) \exp(\gamma_{12}^T Z) =- \alpha_{12}(t|Z) \frac{\phi^{(1)}\{A_{1.}(t|Z)\}}{\phi\{A_{1.}(t|Z)\}} \, ,
\end{equation}  
and
\begin{equation}\label{eq:h013A}
h_{13}(t|Z) =  h_{013}(t) \exp(\gamma_{13}^T Z) = -\alpha_{13}(t|Z) \frac{\phi^{(1)}\{A_{1.}(t|Z)\}}{\phi\{A_{1.}(t|Z)\}} \, .
\end{equation}  
From Eq.s (\ref{eq:stay1}) and (\ref{eq:stay2}) we get $A_{1.}(t|Z) = \psi[\exp\{-H_{1.}(t|Z) \} ]$ and
\begin{eqnarray*}
	\alpha_{1.}(t|Z) = -\psi^{(1)}[\exp\{-H_{1.}(t|Z)\}]\exp\{-H_{1.}(t|Z)\}h_{1.}(t|Z) \, ,
\end{eqnarray*}
where $\psi$ is the inverse Laplace transform, $\alpha_{1.}(t|Z)=\alpha_{12}(t|Z)+\alpha_{12}(t|Z)$  and $h_{1.}(t|Z)=h_{12}(t|Z)+h_{12}(t|Z)$. Also, taking the ratio of (\ref{eq:h012A}) and (\ref{eq:h013A}),
\begin{eqnarray*}
	\frac{\alpha_{12}(t|Z)}{\alpha_{13}(t|Z)} = \frac{h_{12}(t|Z)}{h_{13}(t|Z)} \equiv D(t|Z) \, .
\end{eqnarray*}
Then, letting $\Upsilon(t|Z) = -\psi^{(1)}[\exp\{-H_{1.}(t|Z)\}]\exp\{-H_{1.}(t|Z)\}$, and the above yields 
$$
\alpha_{12}(t|Z) = D(t|Z) \alpha_{13}(t|Z) = D(t|Z) \{\Upsilon(t|Z) h_{1.}(t|Z) - \alpha_{12}(t|Z) \} \, ,
$$
and
$$
\alpha_{12}(t|Z) = \frac{D(t|Z)}{1+D(t|Z)} \Upsilon(t|Z) h_{1.}(t|Z) \, .
$$
Finally, 
\begin{eqnarray}\label{eq:alpha12}
\alpha_{12}(t|Z) &=& \Upsilon(t|Z) h_{12}(t|Z) \nonumber \\
&=& - h_{012}(t) \exp(\gamma_{12}^T Z) \psi^{(1)}[\exp\{-H_{1.}(t|Z)\}]\exp\{-H_{1.}(t|Z)\} 
\end{eqnarray}
and similarly,
\begin{eqnarray}\label{eq:alpha13}
\alpha_{13}(t|Z) &=& \Upsilon(t|Z) h_{13}(t|Z) \nonumber \\
&=& - h_{013}(t) \exp(\gamma_{13}^T Z) \psi^{(1)}[\exp\{-H_{1.}(t|Z)\}]\exp\{-H_{1.}(t|Z)\}  \, .
\end{eqnarray}
Now,  for $t>t_1>0$, the hazard of transition from disease to death is derived by   
\begin{eqnarray}
&&\Pr(2 \rightarrow 3 \in [t,t+\Delta]|T_1=t_1,T_2>t_1,Z)  \nonumber \\
&& \,\,\, = \Delta 
\frac{\int \omega^2 \alpha_{12}(t_1|Z) 
	\alpha_{23}(t|Z) \exp[-\omega\{A_{1.}(t_1|Z)+A_{23}(t|Z)-A_{23}(t_1|Z)\}] dF(\omega)}
{\int \omega \alpha_{12}(t_1|Z) \exp\{-\omega A_{1.}(t_1|Z)\} dF(\omega) } \nonumber \\
&& \,\,\, = - \Delta \alpha_{23}(t|Z) 
\frac{\phi^{(2)} \{A_{1.}(t_1|Z)+A_{23}(t|Z)-A_{23}(t_1|Z)\}} 
{\phi^{(1)}\{A_{1.}(t_1|Z) \}} \nonumber
\end{eqnarray}
and
\begin{eqnarray}
&& \Pr(\mbox{remain in 2 during} \, [t_1,t]|T_1=t_1,T_2>t_1,Z ) \nonumber \\
&& \,\,\,\,\,\, =  
\frac{\int \omega \alpha_{12}(t_1|Z) 
	\exp[-\omega\{A_{1.}(t_1|Z)+A_{23}(t|Z)-A_{23}(t_1|Z)\}] dF(\omega)}
{\int \omega \alpha_{12}(t_1|Z) \exp\{-\omega A_{1.}(t_1|Z)\} dF(\omega) } \nonumber \\
&& \,\,\,\,\,\, =   
\frac{\phi^{(1)} \{A_{1.}(t_1|Z)+A_{23}(t|Z)-A_{23}(t_1|Z)\}} 
{\phi^{(1)}\{A_{1.}(t_1|Z) \}} \, .\nonumber
\end{eqnarray}
Then,  
\begin{eqnarray}
h_{23}(t|t_1,Z) &=& h_{023}(t)\exp(\gamma_{23}^T Z) \nonumber \\
&=& - \alpha_{23}(t|Z)
\frac{\phi^{(2)} \{A_{1.}(t_1|Z)+A_{23}(t|Z)-A_{23}(t_1|Z)\} } 
{\phi^{(1)} \{A_{1.}(t_1|Z)+A_{23}(t|Z)-A_{23}(t_1|Z)\} } \nonumber \\
&=& - \frac{\partial}{\partial t} \log \{-\phi^{(1)}\} \{\mathcal{A}(t_1,t|Z) \} \nonumber
\end{eqnarray}
where $\mathcal{A}(t_1,t|Z)=A_{1.}(t_1|Z)+A_{23}(t|Z)-A_{23}(t_1|Z)$.
Therefore,
\begin{eqnarray}
\mathcal{A}(t_1,t|Z) &=& \xi [\exp\{-H_{23}(t|t_1,Z) \} ] \nonumber \, ,
\end{eqnarray}
where $\xi$ is the inverse of $-\phi^{(1)}$, leading to 
\begin{eqnarray}
A_{23}(t|Z) = \xi [\exp\{-H_{23}(t|t_1,Z) \} ] + A_{23}(t_1|Z) - A_{1.}(t_1|Z) \nonumber \, .
\end{eqnarray}
Finally,
\begin{eqnarray}\label{eq:alpha23}
\alpha_{23}(t|Z) &=& \frac{\partial}{\partial t} \xi [\exp\{-H_{23}(t|t_1,Z) \} ]  \nonumber \\
&=& -\xi^{(1)} [\exp\{-H_{023}(t) \exp(\gamma_{23}^T Z) \} ] \exp\{-H_{023}(t) \exp(\gamma_{23}^T Z) \} \nonumber \\
&& \exp(\gamma_{23}^T Z) h_{023}(t) \, . 
\end{eqnarray}

\section{A short comparison with \cite{gorfine2006prospective}}
Our proposed estimation procedure for the setting of no delayed entry is an extension of \cite{gorfine2006prospective}. It is useful to clarify the similarity and differences between  \cite{gorfine2006prospective} and our current work:  
\begin{itemize}
	\item
	\cite{gorfine2006prospective} considered the standard shared-frailty model of clustered data (e.g. family data) in which the conditional hazard functions given $(Z,\omega)$ are defined by Cox models times the frailty $\omega$, with unspecified baseline hazards. In contrast, here we focus on the marginal hazard functions integrating over $\omega$, and the conditional hazard functions, given $(Z,\omega)$, are dictated by the Cox models imposed on the marginal hazards and the frailty distribution.
	\item 
	\cite{gorfine2006prospective} considered the standard shared-frailty model of clustered data (e.g. family data), in contrast to the illness-death model considered here. Thus, in the setting of Gorfine et al. the number of events within a cluster could be as large as the cluster size. In the current setting, the number of events within a cluster is at most two. 
	\item 
	Under the shared-frailty setting of \cite{gorfine2006prospective}, the estimation procedure uses one $\sigma-$algebra for the stochastic intensity processes. In our case, since death after having the disease can occur only after having the disease, two  $\sigma-$algebras are required.
\end{itemize}

\section{Details of the likelihood function}
It is assumed that conditional on $Z_i$ and $\omega_i$, the censoring times are independent of the failure times and non-informative for $\omega_i$ and all the other parameters in the models. In addition, the frailty variate $\omega_i$ is assumed to be independent of $Z_i$. Then, the likelihood function is proportional to $L(\gamma,\theta,H_0) = \prod_{i=1}^n L_i$, where
\begin{eqnarray} \label{eq:likelihood}
L_i
&=&  \int f(V_i,\delta_{1i},\delta_{2i},W_i,\delta_{3i}|Z_i,\omega)dF(\omega)  \nonumber \\
&=& \int f(V_i,\delta_{1i},\delta_{2i}|Z_i,\omega)f(W_i,\delta_{3i}|V_i,W_i>V_i,\delta_{1i}=1,Z_i,\omega)^{\delta_{1i}}dF(\omega) \nonumber \\
&\propto& \int \{h^c_{12}(V_i|Z_{i},\omega) \}^{\delta_{1i}}
\{h^c_{13}(V_i|Z_{i},\omega) \}^{\delta_{2i}}
\{h^c_{23}(W_i|V_i,Z_{i},\omega) \}^{\delta_{3i}} \nonumber \\
& & \exp [ -\omega \{A_{12}(V_i|Z_i) + A_{13}(V_i|Z_{i}) +\delta_{1i}A_{23}(W_i|Z_{i})-
\delta_{1i}A_{23}(V_i|Z_{i}) \}]dF(\omega) \nonumber \\
&=& \{h_{012}(V_i) \alpha^*_{12}(V_i|Z_{i}) \}^{\delta_{1i}}
\{h_{013}(V_i) \alpha^*_{13}(V_i|Z_{i})\}^{\delta_{2i}}
\{h_{023}(W_i) \alpha^*_{23}(W_i|Z_{i})\}^{\delta_{3i}} \nonumber \\ 
& & \int \omega^{\delta_{.i}} \exp [ -\omega \{A_{12}(V_i|Z_{i}) + A_{13}(V_i|Z_{i}) +\delta_{1i}A_{23}(W_i|Z_{i})- \delta_{1i}A_{23}(V_i|Z_{i}) \}]dF(\omega) \nonumber \\  
&=&
\{h_{012}(V_i) \alpha^*_{12}(V_i|Z_{i}) \}^{\delta_{1i}}
\{h_{013}(V_i) \alpha^*_{13}(V_i|Z_{i})\}^{\delta_{2i}}
\{h_{023}(W_i) \alpha^*_{23}(W_i|Z_{i})\}^{\delta_{3i}} \nonumber \\
& & 
(-1)^{\delta_{.i}} \phi^{(\delta_{.i})}(s)\vert_{s=s_i} \, ,
\end{eqnarray}
$\delta_{.i}=\sum_{j=1}^3 \delta_{ji}$, and
\begin{eqnarray}\label{eq:si}
s_i=A_{12}(V_i|Z_{i})+A_{13}(V_i|Z_{i})+\delta_{1i}A_{23}(W_i|Z_{i})-\delta_{1i}A_{23}(V_i|Z_{i}) \, .
\end{eqnarray}
The above likelihood consists of the fact that $h^c_{23}(t|t_1,Z,\omega)$ is defined on the restricted support $t>t_1>0$, so the respective density function is a truncated density at $t_1$.

\section{Details of $\widehat{A}_{jk}$ under gamma frailty}
The estimators of ${A}_{1k}$, $k=2,3$, under the gamma frailty settings, are derived in the following manner:
\begin{align*}
\widehat{A}_{1k}(t) &= \int_0^t \widehat{h}_{01k}(s)\widehat{\alpha}^{*}_{1k}(s|Z)ds \\
&= \int_0^t \widehat{h}_{01k}(s)\exp(\widehat{\gamma}^T_{1k}Z + \widehat{\theta} \widehat{H}_{1.}(s|Z))ds  \\
&= \sum_{i=0}^{M_{1.}(t)-1} \int_{t_{1.(i)}}^{t_{1.(i+1)}} \widehat{h}_{01k}(s)\exp(\widehat{\gamma}^T_{1k}Z + 
\widehat{\theta} \widehat{H}_{1.}(s|Z))ds \\
&+ \int_{t_{M_{1.(t)}}}^t \widehat{h}_{01k}(s)\exp(\widehat{\gamma}^T_{jk}Z + \widehat{\theta} \widehat{H}_{1.}(s|Z))ds
\end{align*}
where $M_{1.}(t)$ is the ordinal number of the largest observed failure time of the joined times $t_{12}$ and $t_{13}$, which is still smaller than $t$, and $t_{1.(i)}$ is the $i$'th ordered observed time of the joined times.
Now, the function $\widehat{H}_{1.}(t|Z)$ is constant between every pair of consecutive observed failure times, and the function $h_{01k}(t)$ is estimated by the slope of the line connecting two adjacent values of $\widehat{H}_{01k}$. 
As a result, the entire integrand within each interval is constant, so 
$\widehat{A}_{1k}(t)$ is practically calculated as a sum of constants, each multiplied by its respective interval length.

${A}_{23}(t)$ is estimated in a similar fashion, but only times $t_{23}$ are taken into account for the partitioning. 

It should be noted that $A_{jk}$ is estimable just up to the last observed failure time of type $jk$, denoted by $\tau_{jk}$, so that 
$\widehat{A}_{jk}(t)$ is estimated at time $t \wedge \tau_{jk}$.

\section{Assumptions and main steps of the proof of Theorem 1}
The following are the required assumptions for Theorem 1.
\begin{enumerate}
	\item
	The vectors $(V_i,\delta_{1i},\delta_{2i},W_i,\delta_{3i},Z_i)$, $i=1,\dots,n$, are independent and identically distributed.
	\item
	Given the time-independent vector of covariates $Z_i$ and the frailty variate $\omega_i$, the censoring is independent and non-informative of $\omega_i$, $\gamma$ and $H$. In addition, $\omega_i$ is independent of $Z_i$.
	\item
	There exists a finite maximum follow-up time $\tau>0$ such that $E\{ Y_{i(j)}(\tau)\} > 0$ for all $i=1,\ldots,n$ and $j=1,2$.
	\item
	The covariates' vector $Z_i$ is bounded.
	\item 
	The parameter $\mu$ lies in a compact subset containing an open neighborhood of $\mu^o$.
	\item
	The baseline hazard functions $h_{0jk}$, $jk=12,13,23$, are bounded over $[0,\tau]$ by some fixed constant.
	\item 
	The derivative of the frailty density function, with respect to the frailty parameter, is absolutely integrable.
	\item 
	For each subject, there is a positive probability of having the disease before death.
	\item
	Let $U(\mu,\widehat{H}_0)$ be the derivative of $L(\mu,\widehat{H}_0)$ with respect to $\mu$. 
	The matrix $(\partial / \partial \mu) U(\mu,\widehat{H})|_{\mu=\mu^o}$ is invertible with probability going to 1 and $n$ goes to infinity.
\end{enumerate}
Our proposed estimation procedure for the setting of no ascertainment is an extension of \cite{gorfine2006prospective}. The asymptotic theory of the estimators of \cite{gorfine2006prospective}
is provided in details by \cite{zucker2008pseudo}. Hence the proof of Theorem 1 is based on \cite{zucker2008pseudo}.

To emphasize that the hazards' estimators are functions of $\mu$, we write $\widehat{H}_{0jk}(t)$ also as $\widehat{H}_{0jk}(t,\mu)$, $jk=12,13,23$, when needed. Almost sure consistency of $\widehat{\mu}$ and $\widehat{H}_{0jk}$, $jk=12,13,23$, is based on the following results and the consistency theorem of \cite{foutz1977unique}:
\begin{itemize}
	\item[(i)] For $jk=12,13,23$, $\widehat{H}_{0jk}(t,\mu)$ converge almost surely to some functions $H_{0jk}(t,\mu)$, uniformly in $t$ and $\mu$.
	\item[(ii)] The estimating function $U(\mu,H_0(\mu))$ converges uniformly in $\mu$ to $E\{U(\mu,H_0(\mu))\}$, where $H_0(\mu)=(\widehat{H}_{012}(\cdot,\mu),\widehat{H}_{013}(\cdot,\mu),\widehat{H}_{023}(\cdot,\mu))$.
\end{itemize}	 

For the asymptotic normality of $\widehat{\mu}$, write
\begin{eqnarray}\label{eq:asymnor1}
0 &=& U(\widehat\mu,\widehat{H}_0(\widehat\mu)) \nonumber  \\ 
&=& U(\mu^o,H_0^o) + \{ U(\mu^o,\widehat{H}_0(\mu^o)) - U(\mu^o,H_0^o)  \} + \{ U(\widehat\mu,\widehat{H}_0(\mu^o))   - U(\mu^o,\widehat{H}_0(\mu^o)) \}
\end{eqnarray}
and study the right-side of (\ref{eq:asymnor1}). Specifically, $U(\mu^o,H_0^o)$ can be written as a sum of independent identically distributed random vectors with mean zero. Then, the central limit theorem yields $n^{1/2}U(\mu^o,H_0^o)$ is asymptotically mean-zero multivariate normal.

The mean-zero asymptotic normality of $U(\mu^o,\widehat{H}_0(\mu^o)) - U(\mu^o,H_0^o)$ is based on some algebra and the following steps: 
\begin{itemize}
	\item[(1)] A Taylor expansion of $U(\mu^o,\widehat{H}_0(\mu^o))$ about $H_0^o$.
	\item[(2)] Approximating $\widehat{H}_{0jk}-H_{0jk}^o$, $jk=12,13,23$, by a martingale representation (See Eq. (26) of \cite{zucker2008pseudo} and the relevant previous derivations for details). 
	\item[(3)] Plugging the martingale representation of (2) into the expansion of (1).	
\end{itemize}

Taking the first order Taylor expansion of $U(\widehat\mu,\widehat{H}_0(\widehat\mu))$ about $\mu^o$ gives
\begin{equation*}
U(\widehat\mu,\widehat{H}_0(\mu^o))   - U(\mu^o,\widehat{H}_0(\mu^o)) = D(\mu^o)(\widehat{\mu}-\mu^o)=o_p(1) \,
\end{equation*}
where $D(\mu)$ is the derivative of $ U(\mu,\widehat{H}_0(\mu))$ with respect to $\mu$.

Combining the above, we conclude that $n^{1/2}(\widehat{\mu}-\mu^o)$ is asymptotically zero-mean normally distributed.
The asymptotic distributions of $\widehat{H}_{0jk}$, $jk=12,13,23$, are based on writing
\begin{equation}\label{eq:asymnor2}
\widehat{H}_{0jk}(t,\widehat{\mu}) - H^o_{0jk}(t) = \{  \widehat{H}_{0jk}(t,{\mu}) - H^o_{0jk}(t)  \}
+ \{  \widehat{H}_{0jk}(t,\widehat{\mu}) - \widehat{H}_{0jk}(t,{\mu})  \}
\end{equation}
The weak converges and tightness of $\{  \widehat{H}_{0jk}(t,{\mu}) - H^o_{0jk}(t) \}$ is derived from the martingale representation mention in (2) above. Also, by a Taylor expansion we get
$$
\widehat{H}_{0jk}(t,\widehat{\mu}) - \widehat{H}_{0jk}(t,{\mu}) 
=
W(t,\mu^o)^T (\widehat{\mu} - \mu^o) = o_p(1)
$$
where the limiting function of $W(t,\mu)$ is Lipschitz in $t$.
Also, both terms of the right-hand side of Eq. (\ref{eq:asymnor2}) can be written as sums of independent and identically distributed random variables. Thus, by the central limit theorem we get asymptotic normality and tightness, and thus we conclude that $n^{1/2} \{ \widehat{H}_{0jk}(t,\widehat{\mu}) - H^o_{0jk}(t)  \}$, $jk=12,13,23$, converge to Gaussian processes.

\section{The likelihood adjusted to delayed entry}
The likelihood function based on the observed data given the history up to truncation times $R_i$ $i=1,\ldots,n$, is given by
$L^{LT}(\gamma,\theta,H) = \prod_{i: R_i<V_i} L_i^{LT1} \prod_{i: R_i>V_i} L_i^{LT2}$, where 
\begin{eqnarray*}
	L_i^{LT1} &=& f(V_i,\delta_{1i},\delta_{2i},W_i,\delta_{3i}|Z_i,R_i,V_i>R_i) \nonumber \\
	&=& \frac{f(V_i,\delta_{1i},\delta_{2i},W_i,\delta_{3i}|Z_i,R_i)}{f(V_i>R_i|Z_i,R_i)} \nonumber \\
	&=& \frac{\int f(V_i,\delta_{1i},\delta_{2i},W_i,\delta_{3i}|Z_i,R_i,\omega)dF(\omega)}{\int S^c_{12}(R_i|\omega,Z_i)S^c_{13}(R_i|\omega,Z_i)dF(\omega)} \nonumber  \\
	& \propto & 
	\int \{h^c_{12}(V_i|Z_{i},\omega) \}^{\delta_{1i}}
	\{h^c_{13}(V_i|Z_i,\omega) \}^{\delta_{2i}} 
	\{h^c_{23}(W_i|V_i,Z_i,\omega) \}^{\delta_{3i}} \nonumber \\
	& &  \exp[ -w \{ A_{12}(V_i|Z_{i}) + A_{13}(V_i|Z_{i}) 
	+ \delta_{1i} A_{23}(W_i|Z_{i}) - \delta_{1i} A_{23}(V_i|Z_{i}) \} ] dF(\omega)  \nonumber \\
	& & / \int \exp[-w \{ A_{12}(R_i|Z_{i}) + A_{13}(R_i|Z_{i}) \} ] 
	dF(\omega) \nonumber \\
	&=&
	\{h_{012}(V_i) \alpha^*_{12}(V_i|Z_{i}) \}^{\delta_{1i}}
	\{h_{013}(V_i) \alpha^*_{13}(V_i|Z_{i})\}^{\delta_{2i}} 
	\{h_{023}(W_i) \alpha^*_{23}(W_i|Z_{i})\}^{\delta_{3i}} \nonumber \\
	& & 
	(-1)^{\delta_{.i}} \phi^{(\delta_{.i})}(s)\vert_{s=s_{11i}} /\phi(s)\vert_{s=s_{12i}} \, ,
\end{eqnarray*}
where 
$s_{11i} = A_{12}(V_i|Z_{i})+A_{i13}(V_i|Z_{i})+\delta_{1i}A_{23}(W_i|Z_i) - \delta_{1i} A_{23}(V_i|Z_{i})$ and $s_{12i} = A_{12}(R_i|Z_{i})+A_{13}(R_i|Z_{i})$.
Also,
\begin{eqnarray*}
	L_i^{LT2}
	&=& f(V_i,\delta_{1i},\delta_{2i},W_i,\delta_{3i}|Z_i,R_i,V_i,\delta_{1i}=1,\delta_{2i}=0,W_i>R_i) \nonumber \\
	&=& \frac{f(V_i,\delta_{1i}=1,\delta_{2i}=0,W_i,\delta_{3i}|Z_i,R_i)}{f(V_i,\delta_{1i}=1,\delta_{2i}=0,W_i>R_i|Z_i,R_i)} \nonumber \\
	&=& \frac{\int f(V_i,\delta_{1i}=1,\delta_{2i}=0,W_i,\delta_{3i}|Z_i,R_i,\omega)dF(\omega)}{\int f(V_i,\delta_{1i}=1,\delta_{2i}=0,W_i>R_i|Z_i,R_i,\omega)dF(\omega)} \nonumber \\
	&\propto & \int h_{12}^c(V_i|Z_i, \omega) \{h^c_{23}(W_i|V_i,Z_i,\omega) \}^{\delta_{3i}}   \nonumber \\
	& & \exp[ -\omega \{ A_{12}(V_i|Z_{i}) + A_{13}(V_i|Z_{i}) + A_{23}(W_i|Z_{i}) - A_{23}(V_i|Z_{i}) \} ]
	dF(\omega)  \nonumber \\
	& & / \int h_{12}^c(V_i|Z_i ,\omega) 
	\exp[-\omega \{A_{12}(V_i|Z_{i}) + A_{13}(V_i|Z_{i}) + A_{23}(R_i|Z_{i}) - A_{23}(V_i|Z_{i}) \} ] 
	dF(\omega) \nonumber \\
	&=& \{h_{023}(W_i) \alpha^*_{23}(W_i|Z_{i})\}^{\delta_{3i}} 
	(-1)^{1+\delta_{3i}} \phi^{(1+\delta_{3i})}(s)\vert_{s=s_{21i}}  /(-1) \phi^{(1)}(s)\vert_{s=s_{22i}} \, ,
\end{eqnarray*}
where
$s_{21i} = A_{12}(V_i|Z_{i})+A_{13}(V_i|Z_{i})+A_{23}(W_i|Z_{i}) - A_{23}(V_i|Z_{i})$
and
$s_{22i} = A_{12}(V_i|Z_{i})+A_{13}(V_i|Z_{i})+A_{23}(R_i|Z_{i})- A_{23}(V_i|Z_{i})$.

\setlength{\tabcolsep}{5pt}
\begin{table*}[ht]
	\centering
	\caption{{UKbiobank CRC Analysis: Naive Cox Regression Coefficients of E-score \label{escore}} }
	\begin{tabular}{rcccc}
		\hline
		Risk Factor&Coefficient&SE&Z-value&P-value \\
		\hline
		Family History (yes)          & 0.271  & 0.063    & 4.267  & 0.000                 \\
		Height                        & 0.005  & 0.003    & 1.814  & 0.070                 \\
		BMI                           & 0.016  & 0.004    & 3.960  & 0.000                 \\
		smoking (yes)                    & 0.168  & 0.039    & 4.301  & 0.000                 \\
		\multicolumn{5}{l}{Alcohol (reference - non or occasional):}\\
		light frequent drinker        & 0.034  & 0.053    & 0.635  & 0.525                 \\
		very frequent drinker         & 0.093  & 0.048    & 1.949  & 0.051                 \\
		Physical activity (yes)       & -0.010 & 0.041    & -0.255 & 0.799                 \\
		\multicolumn{5}{l}{Education (reference - prefer not to answer):}\\
		lower than high-school     & -0.241 & 0.160    & -1.513 & 0.130                 \\
		high-school                  & -0.240 & 0.158    & -1.516 & 0.130                 \\
		higher vocational education  & -0.243 & 0.163    & -1.490 & 0.136                 \\
		college or university graduate   & -0.304 & 0.159    & -1.910 & 0.056                 \\
		Aspirin use                   & 0.066  & 0.050    & 1.324  & 0.186                 \\
		Ibuprofen drugs use (e.g. Nurofen) & -0.092 & 0.066    & -1.395 & 0.163                 \\
		\multicolumn{5}{l}{Gender and Hormone use (reference - Male):}\\
		Female, no hormones use       & -0.307 & 0.063    & -4.881 & 0.000                 \\
		Female, hormone use           & -0.354 & 0.064    & -5.532 & 0.000 \\             
		\hline
		Quantile-transformed E-score: Mean (SD)\\
		Female diagnosed with CRC & 0.531 (0.289) \\
		Female CRC free & 0.500 (0.289) \\
		Male diagnosed with CRC & 0.588 (0.278) \\
		Male CRC free & 0.500 (0.289) \\
		\hline
	\end{tabular}
\end{table*}

\begin{table*}[ht]
	\centering
	\scriptsize
	\caption{{UKbiobank CRC Analysis: Naive Cox Regression Coefficients of G-score \label{gscore}}}
	\begin{tabular}{rcccccc}
		\hline
		Chromosome and SNP Names & RS Name & 	Position & Coefficient & SE & Z value & P-value  \\
		\hline
		1:22587728\_T/C   & rs72647484  & 22587728  & -0.151 & 0.050    & -3.012 & 0.003                 \\
		1:183081194\_A/C  & rs10911251  & 183081194 & -0.091 & 0.027    & -3.353 & 0.001                 \\
		1:222045446\_G/T  & rs6691170   & 222045446 & 0.068  & 0.028    & 2.464  & 0.014                 \\
		2:192587204\_T/C  & rs11903757  & 192587204 & 0.008  & 0.036    & 0.224  & 0.823                 \\
		2:219154781\_G/A  & rs992157    & 219154781 & 0.071  & 0.027    & 2.608  & 0.009                 \\
		3:37034946\_G/A   & rs1800734   & 37034946  & 0.005  & 0.033    & 0.165  & 0.869                 \\
		3:40924962\_T/A   & rs35360328  & 40924962  & 0.088  & 0.037    & 2.379  & 0.017                 \\
		3:66442435\_C/G   & rs812481    & 66442435  & -0.033 & 0.027    & -1.240 & 0.215                 \\
		3:169492101\_C/T  & rs10936599  & 169492101 & -0.035 & 0.031    & -1.116 & 0.264                 \\
		3:169950156\_T/C  & rs185423955 & 169950156 & 1.363  & 0.502    & 2.717  & 0.007                 \\
		4:94943383\_C/T   & rs1370821   & 94943383  & 0.074  & 0.027    & 2.713  & 0.007                 \\
		4:149748994\_T/C  & rs60745952  & 149748994 & 0.000  & 0.038    & -0.005 & 0.996                 \\
		4:163333405\_T/A  & rs35509282  & 163333405 & -0.022 & 0.043    & -0.507 & 0.612                 \\
		5:1286516\_C/A    & rs2736100   & 1286516   & 0.017  & 0.038    & 0.439  & 0.661                 \\
		5:1296486\_A/G    & rs2735940   & 1296486   & 0.085  & 0.038    & 2.234  & 0.026                 \\
		5:40282106\_G/C   & rs1445012   & 40282106  & 0.128  & 0.029    & 4.406  & 0.000                 \\
		5:96133795\_G/A   & rs142227741 & 96133795  & 0.049  & 0.170    & 0.286  & 0.775                 \\
		5:112175211\_T/A  & rs1801155   & 112175211 & -0.142 & 0.627    & -0.227 & 0.820                 \\
		5:134499092\_C/A  & rs647161    & 134499092 & 0.047  & 0.029    & 1.654  & 0.098                 \\
		6:35528204\_T/C   & rs144037597 & 35528204  & -0.064 & 0.042    & -1.526 & 0.127                 \\
		6:36622900\_C/A   & rs1321311   & 36622900  & 0.004  & 0.031    & 0.140  & 0.889                 \\
		6:41692812\_G/A   & rs4711689   & 41692812  & 0.047  & 0.027    & 1.720  & 0.086                 \\
		6:55714314\_C/T   & rs62404968  & 55714314  & -0.038 & 0.031    & -1.218 & 0.223                 \\
		6:117822993\_C/T  & rs4946260   & 117822993 & 0.037  & 0.027    & 1.366  & 0.172                 \\
		6:160840252\_G/T  & rs7758229   & 160840252 & 0.011  & 0.028    & 0.400  & 0.689                 \\
		8:117624093\_T/C  & rs2450115   & 117624093 & -0.035 & 0.040    & -0.877 & 0.380                 \\
		8:117630683\_A/C  & rs16892766  & 117630683 & 0.178  & 0.046    & 3.837  & 0.000                 \\
		8:117647788\_G/A  & rs6469656   & 117647788 & -0.007 & 0.047    & -0.147 & 0.883                 \\
		8:128413305\_G/T  & rs6983267   & 128413305 & -0.187 & 0.027    & -6.952 & 0.000                 \\
		9:6365683\_A/C    & rs719725    & 6365683   & 0.033  & 0.028    & 1.216  & 0.224                 \\
		10:8701219\_G/A   & rs10795668  & 8701219   & -0.127 & 0.029    & -4.346 & 0.000                 \\
		10:16997266\_G/T  & rs10904849  & 16997266  & 0.008  & 0.029    & 0.284  & 0.776                 \\
		10:52645424\_C/T  & rs10994860  & 52645424  & -0.085 & 0.036    & -2.362 & 0.018                 \\
		10:80819132\_A/G  & rs704017    & 80819132  & 0.110  & 0.027    & 4.036  & 0.000                 \\
		10:101345366\_C/T & rs1035209   & 101345366 & 0.090  & 0.033    & 2.729  & 0.006                 \\
		10:104595248\_G/A & rs4919687   & 104595248 & 0.013  & 0.029    & 0.441  & 0.659                 \\
		10:114280702\_T/C & rs12241008  & 114280702 & 0.121  & 0.042    & 2.856  & 0.004                 \\
		10:114726843\_G/A & rs11196172  & 114726843 & 0.088  & 0.038    & 2.305  & 0.021                 \\
		11:61552680\_G/T  & rs174537    & 61552680  & -0.034 & 0.028    & -1.190 & 0.234                 \\
		11:61982418\_G/A  & rs60892987  & 61982418  & 0.009  & 0.033    & 0.265  & 0.791                 \\               
		\hline
	\end{tabular}
\end{table*}

\begin{table*}[ht]
	\centering
	\scriptsize
	\caption{{UKbiobank CRC Analysis: Naive Cox Regression Coefficients of G-score - Continued \label{gscore_con}}}
	\begin{tabular}{rcccccc}
		\hline
		Chromosome and SNP Names & RS Name & 	Position & Coefficient & SE & Z value & P-value  \\
		\hline
		11:74345550\_T/G  & rs3824999   & 74345550  & 0.087  & 0.027    & 3.243  & 0.001                 \\
		11:111171709\_C/A & rs3802842   & 111171709 & -0.090 & 0.029    & -3.132 & 0.002                 \\
		12:4368352\_T/C   & rs10774214  & 4368352   & -0.043 & 0.028    & -1.575 & 0.115                 \\
		12:4388271\_C/T   & rs3217810   & 4388271   & 0.033  & 0.042    & 0.786  & 0.432                 \\
		12:6385727\_C/T   & rs10849432  & 6385727   & 0.044  & 0.043    & 1.013  & 0.311                 \\
		12:6982162\_C/T   & rs11064437  & 6982162   & -0.188 & 0.192    & -0.978 & 0.328                 \\
		12:51155663\_C/T  & rs11169552  & 51155663  & -0.051 & 0.031    & -1.659 & 0.097                 \\
		12:111884608\_T/C & rs3184504   & 111884608 & 0.079  & 0.027    & 2.969  & 0.003                 \\
		12:115888504\_G/A & rs12822984  & 115888504 & 0.029  & 0.027    & 1.059  & 0.290                 \\
		12:117747590\_T/G & rs73208120  & 117747590 & 0.031  & 0.047    & 0.660  & 0.509                 \\
		13:34093518\_C/G  & rs10161980  & 34093518  & -0.008 & 0.028    & -0.290 & 0.771                 \\
		14:54410919\_T/C  & rs4444235   & 54410919  & 0.076  & 0.027    & 2.826  & 0.005                 \\
		14:54560018\_T/C  & rs1957636   & 54560018  & -0.050 & 0.027    & -1.858 & 0.063                 \\
		15:32993111\_C/T  & rs16969681  & 32993111  & 0.179  & 0.044    & 4.025  & 0.000                 \\
		15:33004247\_G/A  & rs11632715  & 33004247  & 0.066  & 0.027    & 2.444  & 0.015                 \\
		16:9297812\_G/A   & rs79900961  & 9297812   & -0.016 & 0.092    & -0.174 & 0.862                 \\
		16:68820946\_G/A  & rs9929218   & 68820946  & -0.071 & 0.030    & -2.368 & 0.018                 \\
		16:86340448\_G/C  & rs2696839   & 86340448  & -0.047 & 0.027    & -1.755 & 0.079                 \\
		16:86695720\_G/C  & rs16941835  & 86695720  & 0.031  & 0.033    & 0.928  & 0.353                 \\
		17:800593\_T/C    & rs12603526  & 800593    & 0.133  & 0.084    & 1.591  & 0.112                 \\
		18:46450976\_A/G  & rs7229639   & 46450976  & 0.018  & 0.047    & 0.390  & 0.697                 \\
		18:46453463\_T/C  & rs4939827   & 46453463  & -0.154 & 0.028    & -5.458 & 0.000                 \\
		19:33532300\_C/T  & rs10411210  & 33532300  & -0.131 & 0.047    & -2.790 & 0.005                 \\
		19:41860296\_A/G  & rs1800469   & 41860296  & 0.118  & 0.030    & 3.922  & 0.000                 \\
		19:46321507\_A/G  & rs56848936  & 46321507  & -0.628 & 0.717    & -0.876 & 0.381                 \\
		20:6404281\_C/A   & rs961253    & 6404281   & 0.026  & 0.028    & 0.934  & 0.350                 \\
		20:6699595\_T/G   & rs4813802   & 6699595   & 0.084  & 0.028    & 3.030  & 0.002                 \\
		20:7812350\_T/C   & rs2423279   & 7812350   & 0.098  & 0.031    & 3.218  & 0.001                 \\
		20:33173883\_C/T  & rs2295444   & 33173883  & -0.030 & 0.027    & -1.104 & 0.270                 \\
		20:47340117\_A/G  & rs6066825   & 47340117  & -0.178 & 0.029    & -6.250 & 0.000                 \\
		20:49057488\_C/T  & rs1810502   & 49057488  & -0.094 & 0.027    & -3.468 & 0.001                 \\
		20:60921044\_T/C  & rs4925386   & 60921044  & 0.114  & 0.029    & 3.910  & 0.000 \\
		\hline
		Quantile-transformed G-score: Mean (SD)\\
		Female diagnosed with CRC &  0.603 (0.280) \\
		Female CRC free & 0.500 (0.289) \\
		Male diagnosed with CRC & 0.602 (0.280) \\
		Male CRC free & 0.500 (0.289) \\
		\hline
	\end{tabular}
\end{table*}

\section{Additional simulation results - no delayed entry}
Tables \ref{sim_noLT1}--\ref{sim_noLT2} summarize the empirical means and standard deviations of $\widehat{\theta}$, the regression coefficients, and the baseline hazard functions at certain time points, for the simple cohort setting in which observations are followed since birth (i.e. no delayed entry). To save computation time, estimated standard error and empirical coverage rates are not included in the setting of no delayed entry, but included in the setting with delayed entry, presented in the main text.

\section{Illness-death model while $T_1$ and $T_2$ are independent given $Z$}

The partial likelihoods for estimating $\gamma_{12}$, $\gamma_{13}$ and $\gamma_{23}$ while conditioning on the entire observed history up to the recruitment age $R_i$ are given by
$$
\prod_{i=1}^n
\left\{
\frac{\exp(\gamma_{12}^T Z_i)}{\sum_{j=1}^n I(R_j \leq t \leq V_j)\exp(\gamma_{12} Z_j)}
\right\}^{\delta_{1i}I(R_i<V_i)} \, ,
$$ 
$$
\prod_{i=1}^n
\left\{
\frac{\exp(\gamma_{13}^T Z_i)}{\sum_{j=1}^n I(R_j \leq t \leq V_j)\exp(\gamma_{13} Z_j)}
\right\}^{\delta_{2i}I(R_i<V_i)} \, ,
$$ 
and
$$
\prod_{i=1}^n
\left\{
\frac{\exp(\gamma_{23}^T Z_i)}{\sum_{j=1}^n 
	I(R_j \leq t) 
	Y_{j(2)}(V_i)\exp(\gamma_{23} Z_j)}
\right\}^{\delta_{3i}} \, .
$$ 

\begin{table}[]
	\centering
	\spacingset{1.25}
	\footnotesize
	\caption{Simulation results of regression coefficients and dependence parameters for right-censored data with no delayed entry:
		The true parameters values are $\gamma_{12}=(\gamma_{12,1},\gamma_{12,2},\gamma_{12,3},\gamma_{12,4})=
		(2,0.2,0.05,0)$,
		$\gamma_{13}=(\gamma_{13,1},\gamma_{13,2},\gamma_{13,3},\gamma_{13,4})=(0.05,1,0,0)$, and $\gamma_{23}=(\gamma_{23,1},\gamma_{23,2},\gamma_{23,3},\gamma_{23,4})=(1,0,0,0.5)$. Cox is naively corrected for left truncation by $V$ for $jk=23$. To save computation time, estimated standard error and empirical coverage rates are not included in the setting of no delayed entry, but included in the setting with delayed entry, presented in the main text.
		\label{sim_noLT1}}
	\begin{tabular}{clcccccccc}
		$\theta$  &  & $\theta$ & $\gamma_{12,1}$ &	
		$\gamma_{12,2}$ & $\gamma_{12,3}$ &
		$\gamma_{13,1}$ &	
		$\gamma_{13,2}$ & $\gamma_{23,1}$ & $\gamma_{23,4}$ \\
		\hline 
		\multicolumn{9}{c}{Cox Corrected for LT}\\ 
		0 & mean & -   & 2.008 & 0.196 & 0.041 & 0.049 & 1.003 & 0.986 & 0.498  \\ 
		& empirical SD & -  & 0.085 & 0.076 & 0.076 & 0.095 & 0.088 & 0.145 & 0.131 \\ 
		& estimated SE & - & 0.088 & 0.083 & 0.082 & 0.092 & 0.089 & 0.140 & 0.125 \\ 
		& coverage rate & -    & 0.940 & 0.960 & 0.960 & 0.900 & 0.970 & 0.960 & 0.920\\ 
		\multicolumn{9}{c}{The Proposed Approach}\\ 
		0 & mean & 0.050 & 1.987 & 0.194 & 0.038 & 0.043 & 0.993 & 1.042 & 0.495 \\
		& empirical SD & 0.078 & 0.098 & 0.089 & 0.079 & 0.094 & 0.089 & 0.140 & 0.124 \\ 
		%	& bootstrap SE &  &  &  &  &  &  & \\
		%	& coverage rate & &  &  &  &  &  & \\ 
		\multicolumn{9}{c}{Cox Corrected for LT}\\ 
		1&  mean & -&   2.007 & 0.187 & 0.055 & 0.042 & 0.996 & \textbf{0.747} & 0.495  \\  
		& empirical SD & - & 0.084 & 0.090 & 0.088 & 0.093 & 0.085 & 0.149 & 0.153  \\
		& estimated SE &  -&  0.088 & 0.083 & 0.082 & 0.092 & 0.089 & 0.163 & 0.148    \\ 
		& coverage rate &  - &  0.960 & 0.900 & 0.930 & 0.950 & 0.960 & \textbf{0.660} & 0.910   \\ 
		\multicolumn{9}{c}{The Proposed Approach}\\ 
		1& mean & 1.073 & 1.981 & 0.189 & 0.041 & 0.057 & 0.972 & 1.033 & 0.504 \\  
		& empirical SD & 0.131 & 0.090 & 0.079 & 0.083 & 0.091 & 0.087 & 0.153 & 0.146 \\
		%	& estimated SE &  \\ 
		%	& coverage rate &    \\ 
		\multicolumn{9}{c}{Cox Corrected for LT}\\ 
		2  & mean & -& 1.995 & 0.203 & 0.057 & 0.046 & 1.002 & \textbf{0.628} & 0.522 \\ 
		& empirical SD & - & 0.078 & 0.080 & 0.073 & 0.099 & 0.095 & 0.172 & 0.151 \\ 
		& estimated SE & - &0.088 & 0.083 & 0.082 & 0.092 & 0.089 & 0.187 & 0.171 \\ 
		& coverage rate & -  & 0.970 & 0.940 & 0.980 & 0.930 & 0.920 &\textbf{0.440} & 0.940  \\ 
		\multicolumn{9}{c}{The Proposed Approach}\\ 
		2	&  mean & 2.087 & 1.969 & 0.180 & 0.034 & 0.041 & 0.991 & 1.037 & 0.503 \\  
		&  empirical SD & 0.188 & 0.092 & 0.075 & 0.072 & 0.087 & 0.086 & 0.181 & 0.150 \\  
		%	&  bootstrap SE &    \\
		%	&  coverage rate &   \\
		\hline
	\end{tabular}
\end{table}

\begin{table}[]
	\centering
	\spacingset{1}
	\footnotesize 
	\caption{Simulation results of baseline hazard function estimators for right-censored data with no delayed entry: baseline hazard functions at 
		$t=0.1,0.2,0.3,0.4,0.5,0.6$.
		Cox is naively corrected for left truncation by $V$ for $jk=23$. ESD - empirical SD, CR - empirical coverage rate. To save computation time, estimated standard error and empirical coverage rates are not included in the setting of no delayed entry, but included in the setting with delayed entry, presented in the main text.		\label{sim_noLT2}}
	\begin{tabular}{clcccccccccccc}
		$\theta$  &   & 0.1 & 0.2 & 0.3 & 0.4 &	 0.5 & 0.6 & 0.1 & 0.2 & 0.3 & 0.4 & 0.5 & 0.6\\
		\hline 
		& &	\multicolumn{6}{c}{Cox corrected for LT} & \multicolumn{6}{c}{The Proposed Approach}\\
		0 &    $H_{012}(t)$ & 0.050 & 0.150 & 0.250 & 0.350 & 0.450 & 0.550 & 0.050 & 0.150 & 0.250 & 0.350 & 0.450 & 0.550 \\ 
		& mean  &   0.050 & 0.151 & 0.252 & 0.353 & 0.453 & 0.551 & 0.051 & 0.153 & 0.256 & 0.358 & 0.457 & 0.558\\
		& ESD   & 0.004 & 0.012 & 0.020 & 0.027 & 0.038 & 0.045 & 0.006 & 0.017 & 0.028 & 0.037 & 0.047 & 0.056 \\  
		& CR &  0.980 & 0.980 & 0.980 & 0.970 & 0.950 & 0.970 & * & * & * & * & * & *\\
		&  $H_{013}(t)$ & 0.075 & 0.225 & 0.425 & 0.625 & 0.825 & 1.025 & 0.075 & 0.225 & 0.425 & 0.625 & 0.825 & 1.025 \\
		& mean  & 0.075 & 0.225 & 0.426 & 0.626 & 0.822 & 1.021 & 0.075 & 0.227 & 0.429 & 0.633 & 0.835 & 1.042\\
		& ESD  & 0.007 & 0.017 & 0.031 & 0.044 & 0.057 & 0.072 & 0.006 & 0.017 & 0.032 & 0.047 & 0.058 & 0.078 \\  
		& CR &  0.930 & 0.940 & 0.970 & 0.970 & 0.960 & 0.970 & * & * & * & * & * & *\\	
		& $H_{023}(t)$ &  0.000 & 0.080 & 0.180 & 0.280 & 0.380 & 0.480 &  0.000 & 0.080 & 0.180 & 0.280 & 0.380 & 0.480 \\ 
		& mean  & - & 0.082 & 0.183 & 0.285 & 0.385 & 0.486 & - & 0.080 & 0.181 & 0.282 & 0.384 & 0.488 \\
		& ESD  & - & 0.012 & 0.023 & 0.036 & 0.047 & 0.059 & - & 0.011 & 0.020 & 0.032 & 0.042 & 0.055 \\  
		& CR & - & 0.920 & 0.940 & 0.950 & 0.940 & 0.930 & * & * & * & * & * & *\\
		1 &    $H_{012}(t)$ &   0.050 & 0.150 & 0.250 & 0.350 & 0.450 & 0.550 & 0.050 & 0.150 & 0.250 & 0.350 & 0.450 & 0.550 \\ 
		& mean  & 0.050 & 0.150 & 0.251 & 0.351 & 0.452 & 0.558 & 0.052 & 0.154 & 0.256 & 0.361 & 0.461 & 0.567\\
		& ESD  & 0.004 & 0.012 & 0.019 & 0.026 & 0.038 & 0.048 & 0.006 & 0.015 & 0.024 & 0.035 & 0.048 & 0.062\\  
		& CR &  0.980 & 0.970 & 0.950 & 0.970 & 0.960 & 0.970  & * & * & * & * & * & *\\
		&  $H_{013}(t)$ & 0.075 & 0.225 & 0.425 & 0.625 & 0.825 & 1.025  & 0.075 & 0.225 & 0.425 & 0.625 & 0.825 & 1.025   \\ 
		& mean  &  0.076 & 0.227 & 0.430 & 0.632 & 0.837 & 1.033& 0.076 & 0.228 & 0.431 & 0.633 & 0.836 & 1.038 \\
		& ESD  & 0.007 & 0.019 & 0.032 & 0.048 & 0.062 & 0.077& 0.006 & 0.017 & 0.033 & 0.046 & 0.064 & 0.090 \\  
		& CR &  0.940 & 0.940 & 0.950 & 0.940 & 0.940 & 0.950 & * & * & * & * & * & *\\ 
		& $H_{023}(t)$ &  0.000 & 0.080 & 0.180 & 0.280 & 0.380 & 0.480 & 0.000 & 0.080 & 0.180 & 0.280 & 0.380 & 0.480 \\ 
		& mean  &  -& \textbf{0.070} & \textbf{0.145} & \textbf{0.217} & \textbf{0.283} & \textbf{0.347} & - & 0.080 & 0.180 & 0.282 & 0.382 & 0.484\\
		& ESD  & - & 0.010 & 0.020 & 0.028 & 0.036 & 0.044 & - & 0.012 & 0.024 & 0.035 & 0.047 & 0.060 \\  
		& CR & -& \textbf{0.770} & \textbf{0.600} & \textbf{0.400}& \textbf{0.250} & \textbf{0.170} & * & * & * & * & * & * \\
		2 &    $H_{012}(t)$ &   0.050 & 0.150 & 0.250 & 0.350 & 0.450 & 0.550  & 0.050 & 0.150 & 0.250 & 0.350 & 0.450 & 0.550 \\ 
		& mean  & 0.050 & 0.151 & 0.249 & 0.348 & 0.447 & 0.546 & 0.052 & 0.156 & 0.261 & 0.364 & 0.469 & 0.572\\
		& ESD  &  0.004 & 0.011 & 0.018 & 0.028 & 0.036 & 0.048 & 0.004 & 0.013 & 0.021 & 0.029 & 0.036 & 0.049\\  
		& CR &  1.000 & 1.000 & 1.000 & 0.980 & 0.950 & 0.950 & * & * & * & * & * & *\\
		&  $H_{013}(t)$  & 0.075 & 0.225 & 0.425 & 0.625 & 0.825 & 1.025 & 0.075 & 0.225 & 0.425 & 0.625 & 0.825 & 1.025 \\
		& mean  & 0.075 & 0.225 & 0.427 & 0.630 & 0.825 & 1.029 & 0.076 & 0.228 & 0.432 & 0.638 & 0.842 & 1.038\\
		& ESD  & 0.007 & 0.018 & 0.033 & 0.045 & 0.058 & 0.074 & 0.006 & 0.017 & 0.026 & 0.046 & 0.064 & 0.078  \\  
		& CR & 0.940 & 0.940 & 0.940 & 0.950 & 0.930 & 0.950 & * & * & * & * & * & *\\
		& $H_{023}(t)$ & 0.000 & 0.080 & 0.180 & 0.280 & 0.380 & 0.480 &  0.000 & 0.080 & 0.180 & 0.280 & 0.380 & 0.480 \\ 
		& mean  & - & \textbf{0.056} &\textbf{ 0.113} & \textbf{0.164} & \textbf{0.211} & \textbf{0.257} & - & 0.080 & 0.181 & 0.280 & 0.382 & 0.484 \\
		& ESD  & -  & 0.010 & 0.019 & 0.026 & 0.033 & 0.039 & - & 0.013 & 0.027 & 0.042 & 0.056 & 0.072\\  
		& CR &  - & \textbf{0.320} & \textbf{0.100} & \textbf{0.020} & \textbf{0.020} & \textbf{0.010} & * & * & * & * & * & *\\
		\hline
	\end{tabular}
\end{table}

\bibliographystyle{biometrika}
\bibliography{literature}

\end{document}